\documentclass[twocolumn,english]{revtex4-2}
\usepackage[T1]{fontenc}
\usepackage[utf8]{inputenc}
\usepackage{color}
\usepackage{babel}
\usepackage{amsmath}
\usepackage{amssymb}
\usepackage{graphicx}
\usepackage[unicode=true,pdfusetitle,
 bookmarks=true,bookmarksnumbered=false,bookmarksopen=false,
 breaklinks=false,pdfborder={0 0 0},pdfborderstyle={},backref=false,colorlinks=true]
 {hyperref}
\hypersetup{
 linkcolor=blue,citecolor=blue,urlcolor=black}
 
\begin{document}
\title{Wrapping corrections for long range spin chains}
\author{Tamas Gombor}
\affiliation{MTA-ELTE “Momentum” Integrable Quantum Dynamics Research Group, Department
of Theoretical Physics, Eötvös Loránd University}
\affiliation{Holographic QFT Group, Wigner Research Centre for Physics, Budapest,
Hungary}
\begin{abstract}
The long range spin chains play an important role in the gauge/string
duality. The aim of this paper is to generalize the recently introduced
transfer matrices of integrable medium range spin chains to long
range models. These transfer matrices define a large set of conserved
charges for every length of the spin chain. These charges agree with
the original definition of long range spin chains for infinite length.
However, our construction works for every length, providing the definition
of integrable finite size long range spin chains whose spectrum already
contains the wrapping corrections.
\end{abstract}
\maketitle

\section{Introduction}

In the early studies of the planar limit of the $\mathcal{N}=4$ super
Yang-Mills theory it turned out that the anomalous dimensions of single-trace
operators can be obtained from the spectrum of an integrable Hamiltonian
with long range interaction. At one-loop the dilatation operator corresponds
to an integrable nearest neighbor interacting model \citep{Minahan:2002ve}.
For higher loops the interaction range increases, more precisely,
the interaction range is $\ell+1$ at $\ell$-loop.

In the region where the spin chain length \emph{J} is bigger than
the loop order (asymptotic region), the Hamiltonian can be written
as a sum of local densities. For these local operators the integrability
condition can be generalized and it was showed that the Hamiltonian
of the $SU(2)$ sector preserves integrability for higher loops \citep{Beisert:2003tq}.
These local Hamiltonians can be diagonalized with the asymptotic Bethe
Ansatz \citep{Beisert:2005wv} and the result can be generalized to
the full $\mathfrak{psu}(2,2|4)$ spectrum \citep{Beisert:2005fw}.
However, this result is correct only in the asymptotic region. In
the region where the spin chain length $J$ is smaller than the loop
order (wrapping region), wrapping corrections appear \citep{Sieg:2005kd}.
So far, it was not clear whether good spin chain toy models, which
mimic the wrapping corrections, could be found, i.e., even if an asymptotic
Hamiltonian was given, we could not define the corresponding finite
size Hamiltonian.

The solution for the wrapping corrections came from holographic duality.
In the string theory side the scaling dimensions correspond to the
energy spectrum of strings which can be described as a 1+1 dimensional
integrable field theory \citep{Beisert:2010jr}. In the field theory
if we know the dispersion relation and the scattering matrix at infinite
volume then we can calculate the finite volume spectrum as well (at
least in principle). The finite size corrections can be obtained
from the thermodynamic Bethe Ansatz \citep{Gromov:2009bc,Arutyunov:2009ur,Bombardelli:2009ns,Bajnok:2010ke}
and it was showed that they agree with the wrapping corrections \citep{Bajnok:2008bm,Bajnok:2009vm,Balog:2010xa}.

Since the asymptotic data of the string theory (dispersion relation,
scattering matrix) completely defines the finite size corrections,
a natural conclusion is, that the asymptotic data on the spin chain
side should also define the wrapping corrections. In other words,
there must be a procedure that gives the finite size Hamiltonians
from the asymptotic ones. The aim of this paper to present such a
method.

Recently, an algebraic framework was developed for integrable medium
range spin chains (interaction range bigger than two but finite) \citep{Gombor:2021nhn}.
This method gives a recipe how to define transfer matrices which are
the generating functions of the conserved quantities, including the
Hamiltonians. An interesting observation is that, this transfer matrix
is well defined even when the length of the spin chain is smaller
than the interaction range therefore generalizing this method to long
range spin chains, we obtain transfer matrices which define the finite
length Hamiltonians even for the lengths where the wrapping corrections
appear.

\section{Preliminaries}

In this section we summarize the definition of the long range spin
chain following \citep{Beisert:2005wv,Bargheer:2009xy} and specify
our goals.

An integrable long range spin chain has a tower of coupling constant
$\lambda$ dependent commuting charges $\mathcal{Q}_{r}(\lambda)\equiv\mathcal{Q}_{r}$
\footnote{In this paper we use the calligraphic letters $\mathcal{Q},\mathcal{H},\check{\mathcal{L}},\check{\mathcal{R}}$
for the $\lambda$ dependent quantities and the normal letters $Q,H,\check{L},\check{R}$
for the $\lambda$ independent matrices in the series expansion. For
the sake of brevity, we do not write out the argument $\lambda$.} which have the following series expansions
\begin{equation}
\mathcal{Q}_{r}=Q_{r}^{(0)}+\sum_{j=1}^{\infty}\lambda^{j}Q_{r}^{(j)},\label{eq:longQ}
\end{equation}
where $r\geq2$ and the $\lambda$ independent operators $Q_{r}^{(\ell)}$
are sum of local operators with range $r+\ell$
\begin{equation}
Q_{r}^{(\ell)}=\sum_{j=-\infty}^{\infty}q_{j}^{r,\ell}=\sum_{j=-\infty}^{\infty}q_{j,j+1,\dots,j+r+\ell-1}^{r,\ell},
\end{equation}
where the local densities $q_{j}^{r,\ell}\equiv q_{j,j+1,\dots,j+r+\ell-1}^{r,\ell}$
act on the sites $j,j+1,\dots,j+r+\ell-1$. The Hamiltonian is the
charge $\mathcal{Q}_{2}$.

It turned out that, for a fixed nearest neighbor model $\mathcal{Q}_{k}(\lambda=0)$,
a large class of integrable deformations exists. The moduli space
is given by four sets of parameters $\alpha_{r}(\lambda),\beta_{r,s}(\lambda),\gamma_{r,s}(\lambda),\epsilon_{k}(\lambda)$.
The last two sets are unphysical parameters and they correspond to
the linear combinations of the charges $\mathcal{Q}_{r}\to\sum\gamma_{r,s}(\lambda)\mathcal{Q}_{s}$
and the similarity transformations
\begin{equation}
\mathcal{Q}_{r}\to e^{\mathcal{X}}\mathcal{Q}_{r}e^{-\mathcal{X}},\quad\mathcal{X}=\sum_{j=-\infty}^{\infty}\sum_{k}\epsilon_{k}(\lambda)X_{j}^{k},\label{eq:eps}
\end{equation}
where $X_{j}^{k}\equiv X_{j,\dots,j+\ell_{k}-1}^{k}$-s are local
operators with range $\ell_{k}$. The remaining parameters are the
physical ones. The $\alpha_{r}$ and $\beta_{r,s}$ appear in the
rapidity map and the scattering phase \citep{Bargheer:2009xy}.

It is clear that the operators $Q_{k}^{(\ell)}$ can also be defined
on finite length \emph{J} for $J\geq\ell+k$. More concretely, the
Hamiltonian $\mathcal{H}$ on size \emph{J} is defined up to order
$\lambda^{J-2}$ (asymptotic region). Our goal is the find an integrability
preserving method which defines the finite volume version of the asymptotic
Hamiltonians even for higher orders than $\lambda^{J-2}$ (wrapping
region).

\section{Medium range to long range \label{sec:Medium-range-to}}

\begin{figure*}[t]
\begin{centering}
\includegraphics[width=0.99\textwidth]{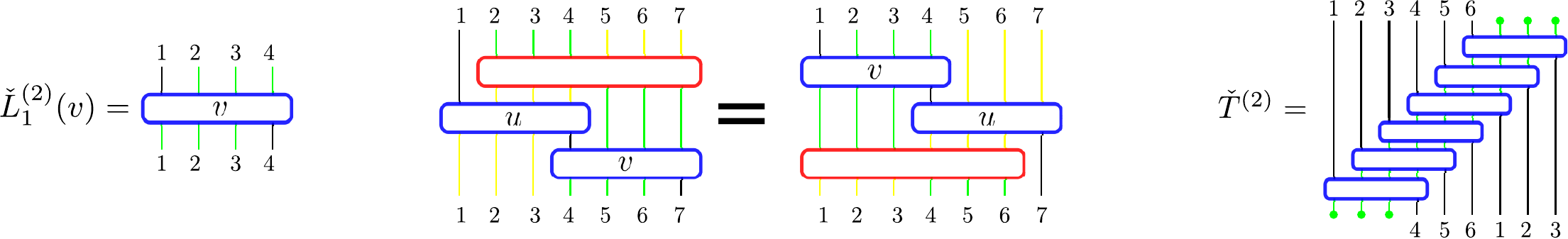}
\par\end{centering}
\caption{Graphical illustration of the Lax-operator, \emph{RLL}-relation and
transfer matrix for $\ell=2$. The left graph shows the Lax operator
$\check{L}_{1}^{(2)}(v)=\check{L}_{1,2,3,4}^{(2)}(v)$. In the middle
we can see the $RLL$-relation where the red box is the \emph{R}-matrix
$\check{R}_{1}^{(2)}(u,v)=\check{R}_{1,2,3,4,5,6}^{(2)}(u,v)$. The
right graph shows the transfer matrix for $J=6$ where the dots on
the incoming and outgoing legs denote the summations for the auxiliary
spaces. The green and yellow lines mark the auxiliary spaces.}

\label{fig:RLL}
\end{figure*}
In this section we generalize the construction of \citep{Gombor:2021nhn}
(the basics appeared first in \citep{Pozsgay_2021}) to obtain transfer
matrices for perturbative long range spin chains \citep{Beisert:2005wv}.
In \citep{Gombor:2021nhn} an algebraic framework was introduced for
integrable spin chains with interaction range $\ell+2$ which is defined
by the Hamiltonian
\begin{equation}
H^{(\ell)}=\sum_{j=1}^{J}h_{j,j+1,\dots,j+\ell+1}=\sum_{j}h_{j}^{(\ell)},
\end{equation}
where $h_{j}^{(\ell)}$ is the Hamiltonian density which acts on the
sites $j,j+1,\dots,j+\ell+1$. We use periodic boundary condition.
The construction of \citep{Gombor:2021nhn} is based on the existence
of the Lax- and the \emph{R}-operators
\begin{align}
\check{L}_{j}^{(\ell)}(u) & =\check{L}_{j,j+1,\dots,j+\ell+1}^{(\ell)}(u)=1+uh_{j}^{(\ell)}+\mathcal{O}(u^{2}),\label{eq:regular}\\
\check{R}_{j}^{(\ell)}(u,v) & =\check{R}_{j,j+1,\dots,j+2\ell+1}^{(\ell)}(u,v),
\end{align}
which satisfy the \emph{RLL}-relation
\begin{equation}
\check{R}_{2}^{(\ell)}(u,v)\check{L}_{1}^{(\ell)}(u)\check{L}_{\ell+2}^{(\ell)}(v)=\check{L}_{1}^{(\ell)}(v)\check{L}_{\ell+2}^{(\ell)}(u)\check{R}_{1}^{(\ell)}(u,v).
\end{equation}
In this letter we chose to write these operators in the ''checked''
form (the \emph{R}-matrix is multiplied by a permutation), which might
be less familiar to some readers \citep{supp_mat}, although it has
the advantage that the Lax-operator has a simpler expansion in the
spectral parameter (\ref{eq:regular}). In the alternative ''unchecked''
convention the quantum and auxiliary spaces are separated. The figure
\ref{fig:RLL} shows graphical presentations of Lax-operators and
\emph{RLL}-relations and the colored legs denote the auxiliary spaces
of the ''unchecked'' convention.

The consequence the $RLL$-relation is that the following transfer
matrix
\begin{equation}
\check{T}^{(\ell)}(u)=\widehat{\mathrm{Tr}}_{J,\ell+1}\left(\check{L}_{J}^{(\ell)}(u)\dots\check{L}_{1}^{(\ell)}(u)\right)\label{eq:transferDef}
\end{equation}
defines commuting quantities $[\check{T}^{(\ell)}(u),\check{T}^{(\ell)}(v)]=0$
\citep{supp_mat}. In (\ref{eq:transferDef}) we defined the twisted
trace operator $\widehat{\mathrm{Tr}}_{J,\ell}$ which acts on an
operator $X$ as
\begin{equation}
\widehat{\mathrm{Tr}}_{J,\ell}\left(X\right)=\mathrm{Tr}_{J+1,\dots,J+\ell}\left(XP_{\ell,J+\ell}P_{\ell-1,J+\ell-1}\dots P_{1,J+1}\right),\label{eq:twTr}
\end{equation}
where $P_{j,k}$ is the permutation operator and $\mathrm{Tr}_{J+1,\dots,J+\ell}$
is the usual trace on the sites $J+1,\dots,J+\ell$. The transfer
matrix generates the local conserved charges
\begin{equation}
Q_{k+1}^{(\ell)}=\frac{\partial^{k}}{\partial u^{k}}\log\check{T}^{(\ell)}(u)\Biggr|_{u=0}.
\end{equation}
The interaction range of $Q_{k}^{(\ell)}$ is $(k-1)\ell+k$ and $H^{(\ell)}=Q_{2}^{(\ell)}$.

Let us turn to the long range spin chains. At first we have to introduce
the coupling constant $\lambda$ dependent truncated operators $\check{\mathcal{L}}_{1}^{(\ell)}(u,\lambda)\equiv\check{\mathcal{L}}_{1}^{(\ell)}(u)$
and $\check{\mathcal{R}}_{1}^{(\ell)}(u,v,\lambda)\equiv\check{\mathcal{R}}_{1}^{(\ell)}(u,v)$
with range $\ell+2$ and $2\ell+2$ as
\begin{align}
\check{\mathcal{L}}_{1}^{(\ell)}(u) & =\check{L}_{1}^{(0)}(u)+\sum_{j=1}^{\ell}\lambda^{j}\check{L}_{1}^{(j)}(u),\label{eq:Lax}\\
\check{\mathcal{R}}_{1}^{(\ell)}(u,v) & =\check{R}_{1}^{(\ell,0)}(u,v)+\sum_{j=1}^{\ell}\lambda^{j}\check{R}_{1}^{(\ell,j)}(u,v),
\end{align}
which satisfy the $RLL$-relation up to order $\mathcal{O}(\lambda^{\ell+1})$:
\begin{equation}
\check{\mathcal{R}}_{2}^{(\ell)}\check{\mathcal{L}}_{1}^{(\ell)}(u)\check{\mathcal{L}}_{\ell+2}^{(\ell)}(v)=\check{\mathcal{L}}_{1}^{(\ell)}(v)\check{\mathcal{L}}_{\ell+2}^{(\ell)}(u)\check{\mathcal{R}}_{1}^{(\ell)}+\mathcal{O}(\lambda^{\ell+1}),\label{eq:trunkRLL}
\end{equation}
where we used the shorthand notation $\check{\mathcal{R}}_{j}^{(\ell)}:=\check{\mathcal{R}}_{j}^{(\ell)}(u,v)$.
We also require that
\begin{equation}
\check{\mathcal{L}}_{1}^{(\ell)}(u)=1+u\mathfrak{h}_{1}^{(\ell)}+\mathcal{O}(u^{2}),\qquad\mathfrak{h}_{1}^{(\ell)}=\sum_{j=0}^{\ell}\lambda^{j}h_{1}^{(j)},\label{eq:Laxexp}
\end{equation}
where $h_{1}^{(j)}$ are $\lambda$ and \emph{u} independent operators
with interaction range $j+2$.

At the first sight we might think that the truncated \emph{RLL}-relation
(\ref{eq:trunkRLL}) and the matrices $\check{R}_{1}^{(\ell,j)}$
are completely independent for every order $\ell$ but it is not true.
It turns out that the equation (\ref{eq:trunkRLL}) up to order $\mathcal{O}(\lambda^{\ell})$
is equivalent with the truncated $RLL$-relation for $\ell-1$ . We
can show that
\begin{equation}
\check{\mathcal{R}}_{1}^{(\ell)}+\mathcal{O}(\lambda^{\ell})=\check{\mathcal{L}}_{\ell+1}^{(\ell-1)}(u)\check{\mathcal{R}}_{1}^{(\ell-1)}\check{\mathcal{J}}_{\ell+1}^{(\ell-1)}(v),\label{eq:Rcon}
\end{equation}
where we defined the perturbative inverse $\check{\mathcal{J}}_{1}^{(\ell-1)}(u)$
as $\check{\mathcal{J}}_{1}^{(\ell-1)}(u)\check{\mathcal{L}}_{1}^{(\ell-1)}(u)=1+\mathcal{O}(\lambda^{\ell})$
\citep{supp_mat}.

The consequence of the equation (\ref{eq:Rcon}) is that the matrices
$\check{R}_{1}^{(\ell,j)}$ are determined by $\check{R}_{1}^{(\ell-1,j)}$
for $j=1,\dots\ell-1$ therefore the full truncated $R$-matrix $\check{\mathcal{R}}_{1}^{(\ell)}$
is completely determined by the matrices $\check{R}_{1}^{(j,j)}$
for $j=1,\dots,\ell$. Fixing the leading order $\check{L}_{1,2}^{(0)},\check{R}_{1,2}^{(0,0)}$
to already known \emph{L}- and \emph{R}-matrices of a nearest neighbor
interacting model, we can calculate the matrices $\check{L}_{1}^{(\ell)}(u),\check{R}_{1}^{(\ell,\ell)}(u,v)$
order by order from the highest order $\lambda^{\ell}$ of the truncated
\emph{RLL}-relation (\ref{eq:trunkRLL}).

As in the medium range case, the transfer matrix
\begin{equation}
\check{\mathcal{T}}^{(\ell)}(u)=\widehat{\mathrm{Tr}}_{J,\ell+1}\left(\check{\mathcal{L}}_{J}^{(\ell)}(u)\dots\check{\mathcal{L}}_{1}^{(\ell)}(u)\right)+\mathcal{O}(\lambda^{\ell+1})\label{eq:transferDef-1}
\end{equation}
defines commuting quantities up to order $\lambda^{\ell+1}$: $[\check{\mathcal{T}}^{(\ell)}(u),\check{\mathcal{T}}^{(\ell)}(v)]=\mathcal{O}(\lambda^{\ell+1})$.
The transfer matrix generates the conserved charges up to order $\mathcal{O}(\lambda^{\ell+1})$
\begin{equation}
\mathcal{Q}_{k+1}^{(\ell)}=\frac{\partial^{k}}{\partial u^{k}}\log\check{\mathcal{T}}^{(\ell)}(u)\Biggr|_{u=0}+\mathcal{O}(\lambda^{\ell+1}),\label{eq:Qdef}
\end{equation}
where
\begin{align}
\mathcal{Q}_{k}^{(\ell)} & =Q_{k}^{(0)}+\sum_{j=1}^{\ell}\lambda^{j}Q_{k}^{(j)}(u)+\mathcal{O}(\lambda^{\ell+1}).
\end{align}
It turns out that the charge $Q_{k}^{(\ell)}$ has interaction range
$\ell+k$.

Since the Lax-operators has the property (\ref{eq:Laxexp}) the Hamiltonian
reads as
\begin{equation}
\mathcal{Q}_{2}^{(\ell)}=\sum_{j=1}^{J}\widehat{\mathrm{Tr}}_{J,\ell+1}\left(\mathfrak{h}_{j}^{(\ell)}\right)+\mathcal{O}(\lambda^{\ell+1}).
\end{equation}
Since $\mathfrak{h}_{j}^{(\ell)}=\sum_{k=0}^{\ell}\lambda^{k}h_{j}^{(k)}$
we obtain that
\begin{equation}
Q_{2}^{(\ell)}=\sum_{j=1}^{J}\widehat{\mathrm{Tr}}_{J,\ell+1}\left(h_{j}^{(\ell)}\right).\label{eq:Q2}
\end{equation}
For the asymptotic region i.e. $J>\ell+1$, we have the identity $\widehat{\mathrm{Tr}}_{J,\ell+1}\left(h_{1}^{(\ell)}\right)=h_{1}^{(\ell)}$
therefore this charge has the usual form $Q_{2}^{(\ell)}=\sum_{j=1}^{J}h_{j}^{(\ell)}$.

Above we showed that the solutions of the \emph{RLL}-relations (\ref{eq:trunkRLL})
define long range charges (\ref{eq:longQ}) in the asymptotic limit.
An important question is that whether the reverse statement is also
true i.e. do there exist Lax-operators for every integrable long range
charges $\mathcal{Q}_{2}$? At this point we do not know the answer.
However I investigated the long range $\mathfrak{gl}(N)$ spin chains
of \citep{Beisert:2005wv} up to order $\mathcal{O}(\lambda^{3})$
. After fixing the unphysical parameters $\gamma(\lambda),\epsilon(\lambda)$,
I found the matrices $\check{\mathcal{L}}_{1}^{(2)}(u)$ which give
the $\mathcal{Q}_{2}^{(2)}$ for every physical parameters $\alpha(\lambda),\beta(\lambda)$
\citep{supp_mat}.

\section{Long range spin chains at the wrapping region\label{sec:Long-range-spin}}

The main advantage of the algebraic construction of the previous section
is that, the transfer matrix is well defined and satisfies $[\check{\mathcal{T}}^{(\ell)}(u),\check{\mathcal{T}}^{(\ell)}(v)]=\mathcal{O}(\lambda^{\ell+1})$
\footnote{The commutation of the transfer matrices follows from the $RLL$-relation
and the derivation is independent from the number of the sites \citep{supp_mat}.} even for $J<\ell+2$ i.e. the wrapping region. So far it was not
clear how to define the Hamiltonian in the wrapping region in an integrability
preserving way but our transfer matrix gives a recipe. We emphasis
that Lax-operator (\ref{eq:Lax}) is an asymptotic, density-like quantity
(since it is defined on an infinite chain and it contains the asymptotic
Hamiltonian density) therefore it describes the elementary physical
interaction. The transfer matrix is a consistent way to ''put'' this
interaction to finite size in a translation invariant and integrability
preserving way.

\begin{figure}[t]
\begin{centering}
\includegraphics[width=0.85\columnwidth]{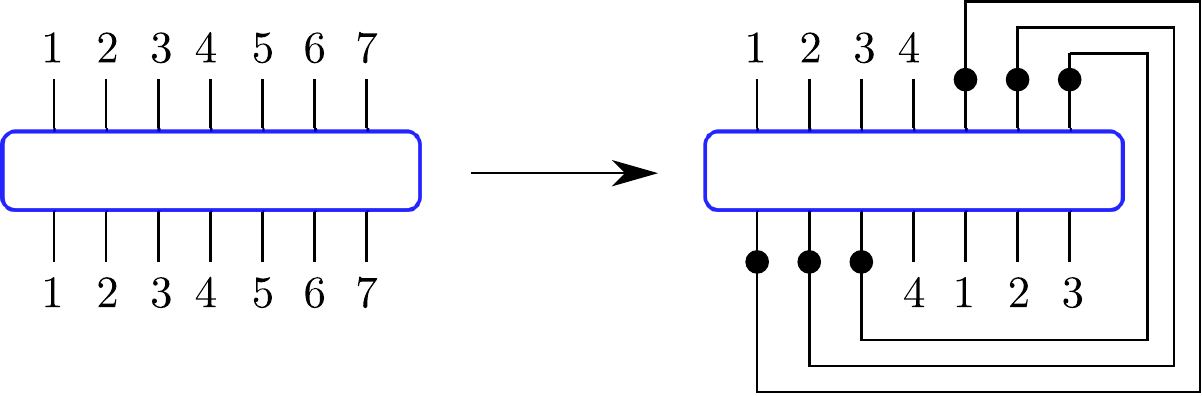}
\par\end{centering}
\caption{In the left there is graph for the seven-site operator $h_{1}^{(5)}=h_{1,2,\dots,7}$.
In the right we can see the wrapped operator $\tilde{h}_{1}^{(5)}$
for $J=4$. The contracted dots denote the summations e.g. we have
to trace out the first incoming and the fifth outgoing legs of the
operator $h_{1,2,\dots,7}$.}

\centering{}\label{fig:wrapping}
\end{figure}

To obtain the integrable Hamiltonian for the wrapping region ($J\leq\ell+1)$,
we only have to use the definition (\ref{eq:Qdef}). We can repeat
the previous calculation up to (\ref{eq:Q2}) i.e., the Hamiltonian
reads as $Q_{2}^{(\ell)}=\sum_{j=1}^{J}\tilde{h}_{j}^{(J,\ell)}$
where we introduced a ''wrapped'' Hamiltonian density (see figure
\ref{fig:wrapping})
\begin{equation}
\tilde{h}_{1}^{(J,\ell)}\equiv\tilde{h}_{12\dots J}^{(J,\ell)}:=\widehat{\mathrm{Tr}}_{J,\ell+2-J}\left(h_{1}^{(\ell)}\right),\label{eq:wrapping}
\end{equation}
 and the periodic boundary condition is prescribed i.e. $\tilde{h}_{j}^{(J,\ell)}=\tilde{h}_{j,j+1,\dots,J,1,2,\dots j-1}^{(J,\ell)}$.
We saw that the twisted trace acts as identity for the asymptotic
region but for the wrapping region it defines a new operator which
''fits'' with the length of the chain.

Let us summarize what we learned from this analysis. Let us take an
asymptotic integrable long range Hamiltonian
\begin{equation}
\mathcal{H}^{\infty}=\sum_{j=-\infty}^{\infty}\mathfrak{h}_{j}^{\infty}=\sum_{j=-\infty}^{\infty}\sum_{\ell=0}^{\infty}\mathcal{X}_{j}^{(\ell)},\label{eq:asymH}
\end{equation}
where $\mathcal{X}_{j}^{(\ell)}\equiv\mathcal{X}_{j,\dots,j+\ell+1}^{(\ell)}$
is a coupling constant dependent operator with interaction range $\ell+2$.
We saw that our method defines a unique integrability preserving Hamiltonian
for every finite length \emph{J} as 
\begin{equation}
\begin{split}\mathcal{H}^{J}= & \sum_{j=1}^{J}\sum_{\ell=0}^{\infty}\tilde{\mathcal{X}}_{j}^{(J,\ell)},\\
\tilde{\mathcal{X}}_{1}^{(J,\ell)}= & \begin{cases}
\mathcal{X}_{1}^{(\ell)}, & \ell+2\leq J,\\
\widehat{\mathrm{Tr}}_{J,\ell+2-J}\left(\mathcal{X}_{1}^{(\ell)}\right), & \ell+2>J.
\end{cases}
\end{split}
\label{eq:finiteH}
\end{equation}

\section{Inozemtsev’s spin chain\label{sec:Inozemtsev=002019s-spin-chain}}

In this section we demonstrate that our finite volume Hamiltonian
is consistent with a naive physical argument. Let us take a long range
interaction and assume that we already know its manifestation for
every length \emph{J}, i.e., for every length \emph{J} we have the
Hamiltonians $\mathcal{H}^{J}$ which correspond to the same physical
interaction. Since we know the Hamiltonians for every length we can
obtain the asymptotic model by the limit
\begin{equation}
\mathcal{H}^{\infty}=\lim_{J\to\infty}\mathcal{H}^{J}.
\end{equation}
A natural requirement is that our procedure (\ref{eq:finiteH}) should
return to the original models $\mathcal{H}^{J}$.

In the following we validate this requirement on the Inozemtsev’s
spin chain \citep{1995JPhA...28L.439I} for which the finite volume
Hamiltonian reads as
\begin{equation}
\mathcal{H}^{J}=\sum_{1\leq j,k\leq J}\left(\wp(k-j)+\frac{2}{\omega}\zeta(\frac{\omega}{2})\right)P_{j,k},\label{eq:inozemtsev}
\end{equation}
where $\wp(z)$, $\zeta(z)$ are the Weierstrass functions defined
on the torus $\mathbb{C}/\mathbb{Z}J+\mathbb{Z}\omega$, $\omega=i\frac{\pi}{\kappa}$
and the local Hilbert spaces are $\mathbb{C}^{N}$. The \emph{J} is
length of the spin chain and $\kappa\in\mathbb{R}$ is the coupling.
In the asymptotic limit we obtain the hyperbolic Inozemtsev’s spin
chain \citep{1995JPhA...28L.439I}
\begin{equation}
\mathcal{H}^{\infty}=\sum_{-\infty<j<k<\infty}V(k-j)P_{j,k},\quad V(j)=\left(\frac{\kappa}{\sinh(j\kappa)}\right)^{2}.
\end{equation}
After a renormalization of the coupling constant $\kappa(\lambda)$,
this Hamiltonian is compatible with the perturbative long range description
\citep{Serban:2004jf}. Let us rewrite the Hamiltonian as
\begin{equation}
\begin{split}\mathcal{H}^{\infty} & =\sum_{-\infty<j<\infty}\mathfrak{h}_{j}^{\infty}=\sum_{\ell=0}^{\infty}\sum_{-\infty<j<\infty}\mathcal{X}_{j}^{(\ell)},\\
\mathcal{X}_{1}^{(\ell)} & =V(\ell+1)P_{1,\ell+2}.
\end{split}
\end{equation}
Now let us apply (\ref{eq:finiteH}) on $\mathcal{X}_{1}^{(\ell)}$.
At first let us wrap the permutations
\begin{equation}
\tilde{P}_{1,k}=\begin{cases}
N, & k\equiv1\mod J\\
P_{1,k_{J}}, & k\equiv2,\dots,J\mod J,
\end{cases}
\end{equation}
where $1<k_{J}\leq J$ and $k_{J}\equiv k\mod J$. Now we can wrap
the full $\mathfrak{h}_{1}^{\infty}=\sum_{\ell=0}^{\infty}\mathcal{X}_{1}^{(\ell)}$
as
\begin{equation}
\mathfrak{h}_{1}^{J}:=\sum_{\ell=0}^{\infty}\mathcal{\tilde{X}}_{1}^{(J,\ell)}=\sum_{1<k\leq J}P_{1,k}\sum_{0\leq l<\infty}V(k+lJ-1)+c,
\end{equation}
where $c=N\sum_{1\leq l<\infty}V(lJ).$ The full finite volume Hamiltonian
is
\begin{equation}
\mathcal{H}^{J}=\sum_{1\leq j\leq J}\mathfrak{h}_{j}^{J}=\sum_{1\leq j<k\leq J}P_{j,k}\sum_{-\infty<l<\infty}V(k-j+lJ)+Jc.
\end{equation}
The infinite sum can be written in the following closed form (23.8.3.
in \citep{olver10})
\begin{equation}
\sum_{-\infty<l<\infty}V(k+lJ)=\wp(k)+\frac{2}{\omega}\zeta(\frac{\omega}{2}).
\end{equation}
Substituting back and dropping the irrelevant identity operator we
just obtained the original Inozemtsev’s Hamiltonian (\ref{eq:inozemtsev}).
We can see that our wrapping method gives the finite volume Inozemtsev’s
spin chain from the infinite volume hyperbolic Inozemtsev’s spin chain
which is an expectation for a consistent wrapping procedure.

\section{Wrapping corrections in ADS/CFT\label{sec:Perspectives-for-ADS/CFT}}

In this section we summarize some properties of the wrapping corrections
in the planar $\mathcal{N}=4$ SYM. We show that our finite volume
Hamiltonians are compatible with these requirements.

\paragraph{Argument 1}

In the string theory side (1+1 dimensional field theory description)
we know that the asymptotic data (dispersion relation and scattering
matrix) defines uniquely the wrapping corrections. This fact is in
agreement with our method which uniquely defines finite size Hamiltonians
(\ref{eq:finiteH}) for a given asymptotic Hamiltonian (\ref{eq:asymH}).

\paragraph{Argument 2}

In the dilation operator of the $\mathcal{N}=4$ SYM there are unfixed
parameters coming from the free choice of the renormalization scheme
\citep{Fiamberti:2007rj,Fiamberti:2008sh}. These are unphysical parameters
which disappear from the spectrum. On the asymptotic level these parameters
correspond to $\epsilon_{k}(\lambda)$ i.e. the global rotations (\ref{eq:eps})
therefore it is clear that they have no effect on the spectrum. The
disappearance on finite volume is a non-trivial condition for the
physical finite size Hamiltonians. It turns out, the spectrum of our
finite volume Hamiltonians is free from $\epsilon_{k}(\lambda)$ as
well \citep{supp_mat}.

\paragraph{Argument 3}

In the asymptotic limit the spectrum of the closed sectors are completely
independent from the full theory. To be more concrete, let us consider
three asymptotic Hamiltonians $\mathcal{H}_{\mathcal{N}=4}^{\infty}$,
$\mathcal{H}_{SU(N)}^{\infty}$ and $\mathcal{H}_{SU(2)}^{\infty}$
which correspond to the $\mathcal{N}=4$ SYM, one of the $SU(N)$
and the $SU(2)$ long range models for which the restriction to the
$SU(2)$ sector are the same i.e.
\[
\mathcal{H}_{\mathcal{N}=4}^{\infty}\Biggr|_{SU(2)}=\mathcal{H}_{SU(N)}^{\infty}\Biggr|_{SU(2)}=\mathcal{H}_{SU(2)}^{\infty}.
\]
Clearly, the spectrum of a closed sector does not know about the full
model in which it is embedded. However, we know that for proper wrapping
corrections we have to consider contributions from the full spectrum
(for Lüsher corrections we have to sum for all virtual particles of
the mirror model \citep{Bajnok:2008bm}) therefore
\[
\mathcal{H}_{\mathcal{N}=4}^{J}\Biggr|_{SU(2)}\neq\mathcal{H}_{SU(2)}^{J}.
\]
This is an important requirement for the definition of the finite
size long range Hamiltonians. Let us take our definition (\ref{eq:finiteH}).
We can see that the wrapped operator $\tilde{\mathcal{X}}_{1}^{(J,\ell)}$
contains a sum for a tensor product of the \emph{full} local Hilbert
spaces! Therefore these wrapped operators, even in the closed sub-sectors,
explicitly depend on the \emph{full asymptotic Hamiltonian} therefore
our definition satisfies that
\[
\mathcal{H}_{SU(N)}^{J}\Biggr|_{SU(2)}\neq\mathcal{H}_{SU(2)}^{J}.
\]

\paragraph{Argument 4}

We also know that, in the wrapping corrections, extra transcendental
numbers appear. For example, let us consider the Konishi operator
(length 4 operator in the $SU(2)$ sector). At four loop, the asymptotic
dilatation operator of the $SU(2)$ sector contains only one transcendental
number $\zeta(3)$ \citep{Fiamberti:2007rj,Fiamberti:2008sh}. However,
the length 4 Hamiltonian at four loop \citep{Fiamberti:2007rj,Fiamberti:2008sh}
contains an extra $\zeta(5)$ compared to the asymptotic Hamiltonian.
We already mentioned that our finite volume Hamiltonian includes a
sum for the full one-site Hilbert space in the wrapping region, therefore
extra transcendental numbers can appear in the finite size Hamiltonians
if the one-site Hilbert space is infinite dimensional, which is the
case for the $\mathcal{N}=4$ SYM. We note that transcendental numbers
appear in the spectrum of higher chargers of nearest neighbor spin
chains with infinite dimensional local Hilbert spaces \citep{Penedones:2008rv}.

\section{Conclusions\label{sec:Conclusions}}

In this paper we generalized the algebraic framework of medium range
spin chains \citep{Gombor:2021nhn} to perturbative long range spin
chains \citep{Beisert:2005wv}. Using this method we were able to
define finite volume Hamiltonians (\ref{eq:finiteH}) for every asymptotic
long range models. We demonstrated that this definition is physically
relevant by showing that our definition is in agreement with several
physical requirements coming form the Inozemtsev’s spin chain and
AdS/CFT.

We saw that our wrapping procedure (\ref{eq:finiteH}) leads to wrapping
corrections with similar properties than what we expect from the $\mathcal{N}=4$
SYM. This is an important result, because so far, the finite size
corrections under simpler conditions could be studied using integrable
field theories. From now on, the wrapping corrections can be also
tested on spin chains, which can be simpler in many ways.

I believe that this result could open up a number of new research
directions. One possible direction is to generalize the integrable
boundary states \citep{Piroli:2017sei,Gombor:2021hmj} for long range
spin chains as well. Combining this to the method of this paper we
could investigate the wrapping corrections of the overlaps between
boundary and Bethe states which describes certain one- and three-point
functions in $\mathcal{N}=4$ SYM \citep{Buhl-Mortensen:2016pxs,Buhl-Mortensen:2017ind,Jiang:2019zig,DeLeeuw:2018cal,deLeeuw:2019ebw,Gombor:2020kgu,Gombor:2020auk}
and ABJM theories \citep{Yang:2021hrl,Kristjansen:2021abc}.

It would be interesting to apply the algebraic Bethe Ansatz, although
it is not clear how this should be done due to the increasing number
of auxiliary spaces. However, there are other ways to diagonalize
the transfer matrices e.g. functional techniques \citep{Gromov:2010kf}
(quantum spectral curve \citep{Gromov:2013pga} for simpler long range
models?) and the separation of variables \citep{Cavaglia:2019pow,Ryan:2018fyo,Maillet:2018bim}.

Another interesting directions would be to give some non-perturbative
definitions of the quantities appearing in this paper (Lax-operators,
transfer matrix); derivation of the Yangian symmetry \citep{Beisert:2007jv}
from our framework; connection for the $T\bar{T}$-deformations of
spin chains \citep{Pozsgay:2019ekd}.

Finally, we need to address a major shortcoming of our method. The
spin chain which appears in the perturbation theory of $\mathcal{N}=4$
SYM is dynamical which means that the Hamiltonian can change the length
of the spin chain. Our method in its present form is not suitable
for describing such models. In the future, we plan to extend the process
to dynamic spin chains, but in the meantime, the non-dynamical Hamiltonians
like (\ref{eq:finiteH}) can serve as good toy models of wrapping
effects.

It is worth to mention that a parallel research is also started on
the topic of long range spin chains \citep{deLeeuw:2022}.\nocite{Gombor:2022ldb}\nocite{Gombor:2022ldb}

\paragraph*{ACKNOWLEDGMENTS}

I thank Balázs Pozsgay, Zoltán Bajnok and László Fehér for the useful
discussions and the NKFIH grant K134946 for support.

\bibliographystyle{apsrev4-2}
\bibliography{ref}

\begin{thebibliography}{45}%
\makeatletter
\providecommand \@ifxundefined [1]{%
 \@ifx{#1\undefined}
}%
\providecommand \@ifnum [1]{%
 \ifnum #1\expandafter \@firstoftwo
 \else \expandafter \@secondoftwo
 \fi
}%
\providecommand \@ifx [1]{%
 \ifx #1\expandafter \@firstoftwo
 \else \expandafter \@secondoftwo
 \fi
}%
\providecommand \natexlab [1]{#1}%
\providecommand \enquote  [1]{``#1''}%
\providecommand \bibnamefont  [1]{#1}%
\providecommand \bibfnamefont [1]{#1}%
\providecommand \citenamefont [1]{#1}%
\providecommand \href@noop [0]{\@secondoftwo}%
\providecommand \href [0]{\begingroup \@sanitize@url \@href}%
\providecommand \@href[1]{\@@startlink{#1}\@@href}%
\providecommand \@@href[1]{\endgroup#1\@@endlink}%
\providecommand \@sanitize@url [0]{\catcode `\\12\catcode `\$12\catcode
  `\&12\catcode `\#12\catcode `\^12\catcode `\_12\catcode `\%12\relax}%
\providecommand \@@startlink[1]{}%
\providecommand \@@endlink[0]{}%
\providecommand \url  [0]{\begingroup\@sanitize@url \@url }%
\providecommand \@url [1]{\endgroup\@href {#1}{\urlprefix }}%
\providecommand \urlprefix  [0]{URL }%
\providecommand \Eprint [0]{\href }%
\providecommand \doibase [0]{https://doi.org/}%
\providecommand \selectlanguage [0]{\@gobble}%
\providecommand \bibinfo  [0]{\@secondoftwo}%
\providecommand \bibfield  [0]{\@secondoftwo}%
\providecommand \translation [1]{[#1]}%
\providecommand \BibitemOpen [0]{}%
\providecommand \bibitemStop [0]{}%
\providecommand \bibitemNoStop [0]{.\EOS\space}%
\providecommand \EOS [0]{\spacefactor3000\relax}%
\providecommand \BibitemShut  [1]{\csname bibitem#1\endcsname}%
\let\auto@bib@innerbib\@empty
\bibitem [{\citenamefont {Minahan}\ and\ \citenamefont
  {Zarembo}(2003)}]{Minahan:2002ve}%
  \BibitemOpen
  \bibfield  {author} {\bibinfo {author} {\bibfnamefont {J.~A.}\ \bibnamefont
  {Minahan}}\ and\ \bibinfo {author} {\bibfnamefont {K.}~\bibnamefont
  {Zarembo}},\ }\href {https://doi.org/10.1088/1126-6708/2003/03/013}
  {\bibfield  {journal} {\bibinfo  {journal} {JHEP}\ }\textbf {\bibinfo
  {volume} {03}},\ \bibinfo {pages} {013}},\ \Eprint
  {https://arxiv.org/abs/hep-th/0212208} {arXiv:hep-th/0212208} \BibitemShut
  {NoStop}%
\bibitem [{\citenamefont {Beisert}\ \emph {et~al.}(2003)\citenamefont
  {Beisert}, \citenamefont {Kristjansen},\ and\ \citenamefont
  {Staudacher}}]{Beisert:2003tq}%
  \BibitemOpen
  \bibfield  {author} {\bibinfo {author} {\bibfnamefont {N.}~\bibnamefont
  {Beisert}}, \bibinfo {author} {\bibfnamefont {C.}~\bibnamefont
  {Kristjansen}},\ and\ \bibinfo {author} {\bibfnamefont {M.}~\bibnamefont
  {Staudacher}},\ }\href {https://doi.org/10.1016/S0550-3213(03)00406-1}
  {\bibfield  {journal} {\bibinfo  {journal} {Nucl. Phys. B}\ }\textbf
  {\bibinfo {volume} {664}},\ \bibinfo {pages} {131} (\bibinfo {year}
  {2003})},\ \Eprint {https://arxiv.org/abs/hep-th/0303060}
  {arXiv:hep-th/0303060} \BibitemShut {NoStop}%
\bibitem [{\citenamefont {Beisert}\ and\ \citenamefont
  {Klose}(2006)}]{Beisert:2005wv}%
  \BibitemOpen
  \bibfield  {author} {\bibinfo {author} {\bibfnamefont {N.}~\bibnamefont
  {Beisert}}\ and\ \bibinfo {author} {\bibfnamefont {T.}~\bibnamefont
  {Klose}},\ }\href {https://doi.org/10.1088/1742-5468/2006/07/P07006}
  {\bibfield  {journal} {\bibinfo  {journal} {J. Stat. Mech.}\ }\textbf
  {\bibinfo {volume} {0607}},\ \bibinfo {pages} {P07006} (\bibinfo {year}
  {2006})},\ \Eprint {https://arxiv.org/abs/hep-th/0510124}
  {arXiv:hep-th/0510124} \BibitemShut {NoStop}%
\bibitem [{\citenamefont {Beisert}\ and\ \citenamefont
  {Staudacher}(2005)}]{Beisert:2005fw}%
  \BibitemOpen
  \bibfield  {author} {\bibinfo {author} {\bibfnamefont {N.}~\bibnamefont
  {Beisert}}\ and\ \bibinfo {author} {\bibfnamefont {M.}~\bibnamefont
  {Staudacher}},\ }\href {https://doi.org/10.1016/j.nuclphysb.2005.06.038}
  {\bibfield  {journal} {\bibinfo  {journal} {Nucl. Phys. B}\ }\textbf
  {\bibinfo {volume} {727}},\ \bibinfo {pages} {1} (\bibinfo {year} {2005})},\
  \Eprint {https://arxiv.org/abs/hep-th/0504190} {arXiv:hep-th/0504190}
  \BibitemShut {NoStop}%
\bibitem [{\citenamefont {Sieg}\ and\ \citenamefont
  {Torrielli}(2005)}]{Sieg:2005kd}%
  \BibitemOpen
  \bibfield  {author} {\bibinfo {author} {\bibfnamefont {C.}~\bibnamefont
  {Sieg}}\ and\ \bibinfo {author} {\bibfnamefont {A.}~\bibnamefont
  {Torrielli}},\ }\href {https://doi.org/10.1016/j.nuclphysb.2005.06.011}
  {\bibfield  {journal} {\bibinfo  {journal} {Nucl. Phys. B}\ }\textbf
  {\bibinfo {volume} {723}},\ \bibinfo {pages} {3} (\bibinfo {year} {2005})},\
  \Eprint {https://arxiv.org/abs/hep-th/0505071} {arXiv:hep-th/0505071}
  \BibitemShut {NoStop}%
\bibitem [{\citenamefont {Beisert}\ \emph {et~al.}(2012)\citenamefont {Beisert}
  \emph {et~al.}}]{Beisert:2010jr}%
  \BibitemOpen
  \bibfield  {author} {\bibinfo {author} {\bibfnamefont {N.}~\bibnamefont
  {Beisert}} \emph {et~al.},\ }\href
  {https://doi.org/10.1007/s11005-011-0529-2} {\bibfield  {journal} {\bibinfo
  {journal} {Lett. Math. Phys.}\ }\textbf {\bibinfo {volume} {99}},\ \bibinfo
  {pages} {3} (\bibinfo {year} {2012})},\ \Eprint
  {https://arxiv.org/abs/1012.3982} {arXiv:1012.3982 [hep-th]} \BibitemShut
  {NoStop}%
\bibitem [{\citenamefont {Gromov}\ \emph {et~al.}(2010)\citenamefont {Gromov},
  \citenamefont {Kazakov}, \citenamefont {Kozak},\ and\ \citenamefont
  {Vieira}}]{Gromov:2009bc}%
  \BibitemOpen
  \bibfield  {author} {\bibinfo {author} {\bibfnamefont {N.}~\bibnamefont
  {Gromov}}, \bibinfo {author} {\bibfnamefont {V.}~\bibnamefont {Kazakov}},
  \bibinfo {author} {\bibfnamefont {A.}~\bibnamefont {Kozak}},\ and\ \bibinfo
  {author} {\bibfnamefont {P.}~\bibnamefont {Vieira}},\ }\href
  {https://doi.org/10.1007/s11005-010-0374-8} {\bibfield  {journal} {\bibinfo
  {journal} {Lett. Math. Phys.}\ }\textbf {\bibinfo {volume} {91}},\ \bibinfo
  {pages} {265} (\bibinfo {year} {2010})},\ \Eprint
  {https://arxiv.org/abs/0902.4458} {arXiv:0902.4458 [hep-th]} \BibitemShut
  {NoStop}%
\bibitem [{\citenamefont {Arutyunov}\ and\ \citenamefont
  {Frolov}(2009)}]{Arutyunov:2009ur}%
  \BibitemOpen
  \bibfield  {author} {\bibinfo {author} {\bibfnamefont {G.}~\bibnamefont
  {Arutyunov}}\ and\ \bibinfo {author} {\bibfnamefont {S.}~\bibnamefont
  {Frolov}},\ }\href {https://doi.org/10.1088/1126-6708/2009/05/068} {\bibfield
   {journal} {\bibinfo  {journal} {JHEP}\ }\textbf {\bibinfo {volume} {05}},\
  \bibinfo {pages} {068}},\ \Eprint {https://arxiv.org/abs/0903.0141}
  {arXiv:0903.0141 [hep-th]} \BibitemShut {NoStop}%
\bibitem [{\citenamefont {Bombardelli}\ \emph {et~al.}(2009)\citenamefont
  {Bombardelli}, \citenamefont {Fioravanti},\ and\ \citenamefont
  {Tateo}}]{Bombardelli:2009ns}%
  \BibitemOpen
  \bibfield  {author} {\bibinfo {author} {\bibfnamefont {D.}~\bibnamefont
  {Bombardelli}}, \bibinfo {author} {\bibfnamefont {D.}~\bibnamefont
  {Fioravanti}},\ and\ \bibinfo {author} {\bibfnamefont {R.}~\bibnamefont
  {Tateo}},\ }\href {https://doi.org/10.1088/1751-8113/42/37/375401} {\bibfield
   {journal} {\bibinfo  {journal} {J. Phys. A}\ }\textbf {\bibinfo {volume}
  {42}},\ \bibinfo {pages} {375401} (\bibinfo {year} {2009})},\ \Eprint
  {https://arxiv.org/abs/0902.3930} {arXiv:0902.3930 [hep-th]} \BibitemShut
  {NoStop}%
\bibitem [{\citenamefont {Bajnok}(2012)}]{Bajnok:2010ke}%
  \BibitemOpen
  \bibfield  {author} {\bibinfo {author} {\bibfnamefont {Z.}~\bibnamefont
  {Bajnok}},\ }\href {https://doi.org/10.1007/s11005-011-0512-y} {\bibfield
  {journal} {\bibinfo  {journal} {Lett. Math. Phys.}\ }\textbf {\bibinfo
  {volume} {99}},\ \bibinfo {pages} {299} (\bibinfo {year} {2012})},\ \Eprint
  {https://arxiv.org/abs/1012.3995} {arXiv:1012.3995 [hep-th]} \BibitemShut
  {NoStop}%
\bibitem [{\citenamefont {Bajnok}\ and\ \citenamefont
  {Janik}(2009)}]{Bajnok:2008bm}%
  \BibitemOpen
  \bibfield  {author} {\bibinfo {author} {\bibfnamefont {Z.}~\bibnamefont
  {Bajnok}}\ and\ \bibinfo {author} {\bibfnamefont {R.~A.}\ \bibnamefont
  {Janik}},\ }\href {https://doi.org/10.1016/j.nuclphysb.2008.08.020}
  {\bibfield  {journal} {\bibinfo  {journal} {Nucl. Phys. B}\ }\textbf
  {\bibinfo {volume} {807}},\ \bibinfo {pages} {625} (\bibinfo {year}
  {2009})},\ \Eprint {https://arxiv.org/abs/0807.0399} {arXiv:0807.0399
  [hep-th]} \BibitemShut {NoStop}%
\bibitem [{\citenamefont {Bajnok}\ \emph {et~al.}(2010)\citenamefont {Bajnok},
  \citenamefont {Hegedus}, \citenamefont {Janik},\ and\ \citenamefont
  {Lukowski}}]{Bajnok:2009vm}%
  \BibitemOpen
  \bibfield  {author} {\bibinfo {author} {\bibfnamefont {Z.}~\bibnamefont
  {Bajnok}}, \bibinfo {author} {\bibfnamefont {A.}~\bibnamefont {Hegedus}},
  \bibinfo {author} {\bibfnamefont {R.~A.}\ \bibnamefont {Janik}},\ and\
  \bibinfo {author} {\bibfnamefont {T.}~\bibnamefont {Lukowski}},\ }\href
  {https://doi.org/10.1016/j.nuclphysb.2009.10.015} {\bibfield  {journal}
  {\bibinfo  {journal} {Nucl. Phys. B}\ }\textbf {\bibinfo {volume} {827}},\
  \bibinfo {pages} {426} (\bibinfo {year} {2010})},\ \Eprint
  {https://arxiv.org/abs/0906.4062} {arXiv:0906.4062 [hep-th]} \BibitemShut
  {NoStop}%
\bibitem [{\citenamefont {Balog}\ and\ \citenamefont
  {Hegedus}(2010)}]{Balog:2010xa}%
  \BibitemOpen
  \bibfield  {author} {\bibinfo {author} {\bibfnamefont {J.}~\bibnamefont
  {Balog}}\ and\ \bibinfo {author} {\bibfnamefont {A.}~\bibnamefont
  {Hegedus}},\ }\href {https://doi.org/10.1007/JHEP06(2010)080} {\bibfield
  {journal} {\bibinfo  {journal} {JHEP}\ }\textbf {\bibinfo {volume} {06}},\
  \bibinfo {pages} {080}},\ \Eprint {https://arxiv.org/abs/1002.4142}
  {arXiv:1002.4142 [hep-th]} \BibitemShut {NoStop}%
\bibitem [{\citenamefont {Gombor}\ and\ \citenamefont
  {Pozsgay}(2021)}]{Gombor:2021nhn}%
  \BibitemOpen
  \bibfield  {author} {\bibinfo {author} {\bibfnamefont {T.}~\bibnamefont
  {Gombor}}\ and\ \bibinfo {author} {\bibfnamefont {B.}~\bibnamefont
  {Pozsgay}},\ }\href {https://doi.org/10.1103/PhysRevE.104.054123} {\bibfield
  {journal} {\bibinfo  {journal} {Phys. Rev. E}\ }\textbf {\bibinfo {volume}
  {104}},\ \bibinfo {pages} {054123} (\bibinfo {year} {2021})},\ \Eprint
  {https://arxiv.org/abs/2108.02053} {arXiv:2108.02053 [nlin.SI]} \BibitemShut
  {NoStop}%
\bibitem [{\citenamefont {Bargheer}\ \emph {et~al.}(2009)\citenamefont
  {Bargheer}, \citenamefont {Beisert},\ and\ \citenamefont
  {Loebbert}}]{Bargheer:2009xy}%
  \BibitemOpen
  \bibfield  {author} {\bibinfo {author} {\bibfnamefont {T.}~\bibnamefont
  {Bargheer}}, \bibinfo {author} {\bibfnamefont {N.}~\bibnamefont {Beisert}},\
  and\ \bibinfo {author} {\bibfnamefont {F.}~\bibnamefont {Loebbert}},\ }\href
  {https://doi.org/10.1088/1751-8113/42/28/285205} {\bibfield  {journal}
  {\bibinfo  {journal} {J. Phys. A}\ }\textbf {\bibinfo {volume} {42}},\
  \bibinfo {pages} {285205} (\bibinfo {year} {2009})},\ \Eprint
  {https://arxiv.org/abs/0902.0956} {arXiv:0902.0956 [hep-th]} \BibitemShut
  {NoStop}%
\bibitem [{Note1()}]{Note1}%
  \BibitemOpen
  \bibinfo {note} {In this paper we use the calligraphic letters $\protect
  \mathcal {Q},\protect \mathcal {H},\protect \check {\protect \mathcal
  {L}},\protect \check {\protect \mathcal {R}}$ for the $\lambda $ dependent
  quantities and the normal letters $Q,H,\protect \check {L},\protect \check
  {R}$ for the $\lambda $ independent matrices in the series expansion. For the
  sake of brevity, we do not write out the argument $\lambda $.}\BibitemShut
  {Stop}%
\bibitem [{\citenamefont {Pozsgay}(2021)}]{Pozsgay_2021}%
  \BibitemOpen
  \bibfield  {author} {\bibinfo {author} {\bibfnamefont {B.}~\bibnamefont
  {Pozsgay}},\ }\href {https://doi.org/10.1088/1751-8121/ac1dbf} {\bibfield
  {journal} {\bibinfo  {journal} {J. Phys. A}\ }\textbf {\bibinfo {volume}
  {54}},\ \bibinfo {pages} {384001} (\bibinfo {year} {2021})}\BibitemShut
  {NoStop}%
\bibitem [{sup()}]{supp_mat}%
  \BibitemOpen
  \href@noop {} {\bibinfo {title} {For the details see the supplemental
  material, which includes refs. [3,14,45].}}\BibitemShut {Stop}%
\bibitem [{Note2()}]{Note2}%
  \BibitemOpen
  \bibinfo {note} {The commutation of the transfer matrices follows from the
  $RLL$-relation and the derivation is independent from the number of the sites
  \protect \citep {supp_mat}.}\BibitemShut {Stop}%
\bibitem [{\citenamefont {{Inozemtsev}}(1995)}]{1995JPhA...28L.439I}%
  \BibitemOpen
  \bibfield  {author} {\bibinfo {author} {\bibfnamefont {V.~I.}\ \bibnamefont
  {{Inozemtsev}}},\ }\href {https://doi.org/10.1088/0305-4470/28/16/004}
  {\bibfield  {journal} {\bibinfo  {journal} {Journal of Physics A Mathematical
  General}\ }\textbf {\bibinfo {volume} {28}},\ \bibinfo {pages} {L439}
  (\bibinfo {year} {1995})},\ \Eprint {https://arxiv.org/abs/cond-mat/9504096}
  {arXiv:cond-mat/9504096 [cond-mat]} \BibitemShut {NoStop}%
\bibitem [{\citenamefont {Serban}\ and\ \citenamefont
  {Staudacher}(2004)}]{Serban:2004jf}%
  \BibitemOpen
  \bibfield  {author} {\bibinfo {author} {\bibfnamefont {D.}~\bibnamefont
  {Serban}}\ and\ \bibinfo {author} {\bibfnamefont {M.}~\bibnamefont
  {Staudacher}},\ }\href {https://doi.org/10.1088/1126-6708/2004/06/001}
  {\bibfield  {journal} {\bibinfo  {journal} {JHEP}\ }\textbf {\bibinfo
  {volume} {06}},\ \bibinfo {pages} {001}},\ \Eprint
  {https://arxiv.org/abs/hep-th/0401057} {arXiv:hep-th/0401057} \BibitemShut
  {NoStop}%
\bibitem [{\citenamefont {Olver}\ \emph {et~al.}(2010)\citenamefont {Olver}, ,
  \citenamefont {Lozier}, \citenamefont {Boisvert},\ and\ \citenamefont
  {Clark}}]{olver10}%
  \BibitemOpen
  \bibfield  {author} {\bibinfo {author} {\bibfnamefont {F.~W.~J.}\
  \bibnamefont {Olver}}, , \bibinfo {author} {\bibfnamefont {D.~W.}\
  \bibnamefont {Lozier}}, \bibinfo {author} {\bibfnamefont {R.~F.}\
  \bibnamefont {Boisvert}},\ and\ \bibinfo {author} {\bibfnamefont {C.~W.}\
  \bibnamefont {Clark}},\ }\href@noop {} {\emph {\bibinfo {title} {The {NIST}
  Handbook of Mathematical Functions}}}\ (\bibinfo  {publisher} {Cambridge
  Univ. Press},\ \bibinfo {year} {2010})\BibitemShut {NoStop}%
\bibitem [{\citenamefont {Fiamberti}\ \emph
  {et~al.}(2008{\natexlab{a}})\citenamefont {Fiamberti}, \citenamefont
  {Santambrogio}, \citenamefont {Sieg},\ and\ \citenamefont
  {Zanon}}]{Fiamberti:2007rj}%
  \BibitemOpen
  \bibfield  {author} {\bibinfo {author} {\bibfnamefont {F.}~\bibnamefont
  {Fiamberti}}, \bibinfo {author} {\bibfnamefont {A.}~\bibnamefont
  {Santambrogio}}, \bibinfo {author} {\bibfnamefont {C.}~\bibnamefont {Sieg}},\
  and\ \bibinfo {author} {\bibfnamefont {D.}~\bibnamefont {Zanon}},\ }\href
  {https://doi.org/10.1016/j.physletb.2008.06.061} {\bibfield  {journal}
  {\bibinfo  {journal} {Phys. Lett. B}\ }\textbf {\bibinfo {volume} {666}},\
  \bibinfo {pages} {100} (\bibinfo {year} {2008}{\natexlab{a}})},\ \Eprint
  {https://arxiv.org/abs/0712.3522} {arXiv:0712.3522 [hep-th]} \BibitemShut
  {NoStop}%
\bibitem [{\citenamefont {Fiamberti}\ \emph
  {et~al.}(2008{\natexlab{b}})\citenamefont {Fiamberti}, \citenamefont
  {Santambrogio}, \citenamefont {Sieg},\ and\ \citenamefont
  {Zanon}}]{Fiamberti:2008sh}%
  \BibitemOpen
  \bibfield  {author} {\bibinfo {author} {\bibfnamefont {F.}~\bibnamefont
  {Fiamberti}}, \bibinfo {author} {\bibfnamefont {A.}~\bibnamefont
  {Santambrogio}}, \bibinfo {author} {\bibfnamefont {C.}~\bibnamefont {Sieg}},\
  and\ \bibinfo {author} {\bibfnamefont {D.}~\bibnamefont {Zanon}},\ }\href
  {https://doi.org/10.1016/j.nuclphysb.2008.07.014} {\bibfield  {journal}
  {\bibinfo  {journal} {Nucl. Phys. B}\ }\textbf {\bibinfo {volume} {805}},\
  \bibinfo {pages} {231} (\bibinfo {year} {2008}{\natexlab{b}})},\ \Eprint
  {https://arxiv.org/abs/0806.2095} {arXiv:0806.2095 [hep-th]} \BibitemShut
  {NoStop}%
\bibitem [{\citenamefont {Penedones}\ and\ \citenamefont
  {Vieira}(2008)}]{Penedones:2008rv}%
  \BibitemOpen
  \bibfield  {author} {\bibinfo {author} {\bibfnamefont {J.}~\bibnamefont
  {Penedones}}\ and\ \bibinfo {author} {\bibfnamefont {P.}~\bibnamefont
  {Vieira}},\ }\href {https://doi.org/10.1088/1126-6708/2008/08/020} {\bibfield
   {journal} {\bibinfo  {journal} {JHEP}\ }\textbf {\bibinfo {volume} {08}},\
  \bibinfo {pages} {020}},\ \Eprint {https://arxiv.org/abs/0806.1047}
  {arXiv:0806.1047 [hep-th]} \BibitemShut {NoStop}%
\bibitem [{\citenamefont {Piroli}\ \emph {et~al.}(2017)\citenamefont {Piroli},
  \citenamefont {Pozsgay},\ and\ \citenamefont {Vernier}}]{Piroli:2017sei}%
  \BibitemOpen
  \bibfield  {author} {\bibinfo {author} {\bibfnamefont {L.}~\bibnamefont
  {Piroli}}, \bibinfo {author} {\bibfnamefont {B.}~\bibnamefont {Pozsgay}},\
  and\ \bibinfo {author} {\bibfnamefont {E.}~\bibnamefont {Vernier}},\ }\href
  {https://doi.org/10.1016/j.nuclphysb.2017.10.012} {\bibfield  {journal}
  {\bibinfo  {journal} {Nucl. Phys. B}\ }\textbf {\bibinfo {volume} {925}},\
  \bibinfo {pages} {362} (\bibinfo {year} {2017})},\ \Eprint
  {https://arxiv.org/abs/1709.04796} {arXiv:1709.04796 [cond-mat.stat-mech]}
  \BibitemShut {NoStop}%
\bibitem [{\citenamefont {Gombor}(2022)}]{Gombor:2021hmj}%
  \BibitemOpen
  \bibfield  {author} {\bibinfo {author} {\bibfnamefont {T.}~\bibnamefont
  {Gombor}},\ }\href {https://doi.org/10.1016/j.nuclphysb.2022.115909}
  {\bibfield  {journal} {\bibinfo  {journal} {Nucl. Phys. B}\ }\textbf
  {\bibinfo {volume} {983}},\ \bibinfo {pages} {115909} (\bibinfo {year}
  {2022})},\ \Eprint {https://arxiv.org/abs/2110.07960} {arXiv:2110.07960
  [hep-th]} \BibitemShut {NoStop}%
\bibitem [{\citenamefont {Buhl-Mortensen}\ \emph {et~al.}(2016)\citenamefont
  {Buhl-Mortensen}, \citenamefont {de~Leeuw}, \citenamefont {Ipsen},
  \citenamefont {Kristjansen},\ and\ \citenamefont
  {Wilhelm}}]{Buhl-Mortensen:2016pxs}%
  \BibitemOpen
  \bibfield  {author} {\bibinfo {author} {\bibfnamefont {I.}~\bibnamefont
  {Buhl-Mortensen}}, \bibinfo {author} {\bibfnamefont {M.}~\bibnamefont
  {de~Leeuw}}, \bibinfo {author} {\bibfnamefont {A.~C.}\ \bibnamefont {Ipsen}},
  \bibinfo {author} {\bibfnamefont {C.}~\bibnamefont {Kristjansen}},\ and\
  \bibinfo {author} {\bibfnamefont {M.}~\bibnamefont {Wilhelm}},\ }\href
  {https://doi.org/10.1103/PhysRevLett.117.231603} {\bibfield  {journal}
  {\bibinfo  {journal} {Phys. Rev. Lett.}\ }\textbf {\bibinfo {volume} {117}},\
  \bibinfo {pages} {231603} (\bibinfo {year} {2016})},\ \Eprint
  {https://arxiv.org/abs/1606.01886} {arXiv:1606.01886 [hep-th]} \BibitemShut
  {NoStop}%
\bibitem [{\citenamefont {Buhl-Mortensen}\ \emph {et~al.}(2017)\citenamefont
  {Buhl-Mortensen}, \citenamefont {de~Leeuw}, \citenamefont {Ipsen},
  \citenamefont {Kristjansen},\ and\ \citenamefont
  {Wilhelm}}]{Buhl-Mortensen:2017ind}%
  \BibitemOpen
  \bibfield  {author} {\bibinfo {author} {\bibfnamefont {I.}~\bibnamefont
  {Buhl-Mortensen}}, \bibinfo {author} {\bibfnamefont {M.}~\bibnamefont
  {de~Leeuw}}, \bibinfo {author} {\bibfnamefont {A.~C.}\ \bibnamefont {Ipsen}},
  \bibinfo {author} {\bibfnamefont {C.}~\bibnamefont {Kristjansen}},\ and\
  \bibinfo {author} {\bibfnamefont {M.}~\bibnamefont {Wilhelm}},\ }\href
  {https://doi.org/10.1103/PhysRevLett.119.261604} {\bibfield  {journal}
  {\bibinfo  {journal} {Phys. Rev. Lett.}\ }\textbf {\bibinfo {volume} {119}},\
  \bibinfo {pages} {261604} (\bibinfo {year} {2017})},\ \Eprint
  {https://arxiv.org/abs/1704.07386} {arXiv:1704.07386 [hep-th]} \BibitemShut
  {NoStop}%
\bibitem [{\citenamefont {Jiang}\ \emph {et~al.}(2019)\citenamefont {Jiang},
  \citenamefont {Komatsu},\ and\ \citenamefont {Vescovi}}]{Jiang:2019zig}%
  \BibitemOpen
  \bibfield  {author} {\bibinfo {author} {\bibfnamefont {Y.}~\bibnamefont
  {Jiang}}, \bibinfo {author} {\bibfnamefont {S.}~\bibnamefont {Komatsu}},\
  and\ \bibinfo {author} {\bibfnamefont {E.}~\bibnamefont {Vescovi}},\ }\href
  {https://doi.org/10.1103/PhysRevLett.123.191601} {\bibfield  {journal}
  {\bibinfo  {journal} {Phys. Rev. Lett.}\ }\textbf {\bibinfo {volume} {123}},\
  \bibinfo {pages} {191601} (\bibinfo {year} {2019})},\ \Eprint
  {https://arxiv.org/abs/1907.11242} {arXiv:1907.11242 [hep-th]} \BibitemShut
  {NoStop}%
\bibitem [{\citenamefont {De~Leeuw}\ \emph {et~al.}(2018)\citenamefont
  {De~Leeuw}, \citenamefont {Kristjansen},\ and\ \citenamefont
  {Linardopoulos}}]{DeLeeuw:2018cal}%
  \BibitemOpen
  \bibfield  {author} {\bibinfo {author} {\bibfnamefont {M.}~\bibnamefont
  {De~Leeuw}}, \bibinfo {author} {\bibfnamefont {C.}~\bibnamefont
  {Kristjansen}},\ and\ \bibinfo {author} {\bibfnamefont {G.}~\bibnamefont
  {Linardopoulos}},\ }\href {https://doi.org/10.1016/j.physletb.2018.03.083}
  {\bibfield  {journal} {\bibinfo  {journal} {Phys. Lett. B}\ }\textbf
  {\bibinfo {volume} {781}},\ \bibinfo {pages} {238} (\bibinfo {year}
  {2018})},\ \Eprint {https://arxiv.org/abs/1802.01598} {arXiv:1802.01598
  [hep-th]} \BibitemShut {NoStop}%
\bibitem [{\citenamefont {De~Leeuw}\ \emph {et~al.}(2020)\citenamefont
  {De~Leeuw}, \citenamefont {Gombor}, \citenamefont {Kristjansen},
  \citenamefont {Linardopoulos},\ and\ \citenamefont
  {Pozsgay}}]{deLeeuw:2019ebw}%
  \BibitemOpen
  \bibfield  {author} {\bibinfo {author} {\bibfnamefont {M.}~\bibnamefont
  {De~Leeuw}}, \bibinfo {author} {\bibfnamefont {T.}~\bibnamefont {Gombor}},
  \bibinfo {author} {\bibfnamefont {C.}~\bibnamefont {Kristjansen}}, \bibinfo
  {author} {\bibfnamefont {G.}~\bibnamefont {Linardopoulos}},\ and\ \bibinfo
  {author} {\bibfnamefont {B.}~\bibnamefont {Pozsgay}},\ }\href
  {https://doi.org/10.1007/JHEP01(2020)176} {\bibfield  {journal} {\bibinfo
  {journal} {JHEP}\ }\textbf {\bibinfo {volume} {01}},\ \bibinfo {pages}
  {176}},\ \Eprint {https://arxiv.org/abs/1912.09338} {arXiv:1912.09338
  [hep-th]} \BibitemShut {NoStop}%
\bibitem [{\citenamefont {Gombor}\ and\ \citenamefont
  {Bajnok}(2020)}]{Gombor:2020kgu}%
  \BibitemOpen
  \bibfield  {author} {\bibinfo {author} {\bibfnamefont {T.}~\bibnamefont
  {Gombor}}\ and\ \bibinfo {author} {\bibfnamefont {Z.}~\bibnamefont
  {Bajnok}},\ }\href {https://doi.org/10.1007/JHEP10(2020)123} {\bibfield
  {journal} {\bibinfo  {journal} {JHEP}\ }\textbf {\bibinfo {volume} {10}},\
  \bibinfo {pages} {123}},\ \Eprint {https://arxiv.org/abs/2004.11329}
  {arXiv:2004.11329 [hep-th]} \BibitemShut {NoStop}%
\bibitem [{\citenamefont {Gombor}\ and\ \citenamefont
  {Bajnok}(2021)}]{Gombor:2020auk}%
  \BibitemOpen
  \bibfield  {author} {\bibinfo {author} {\bibfnamefont {T.}~\bibnamefont
  {Gombor}}\ and\ \bibinfo {author} {\bibfnamefont {Z.}~\bibnamefont
  {Bajnok}},\ }\href {https://doi.org/10.1007/JHEP03(2021)222} {\bibfield
  {journal} {\bibinfo  {journal} {JHEP}\ }\textbf {\bibinfo {volume} {03}},\
  \bibinfo {pages} {222}},\ \Eprint {https://arxiv.org/abs/2006.16151}
  {arXiv:2006.16151 [hep-th]} \BibitemShut {NoStop}%
\bibitem [{\citenamefont {Yang}\ \emph {et~al.}(2022)\citenamefont {Yang},
  \citenamefont {Jiang}, \citenamefont {Komatsu},\ and\ \citenamefont
  {Wu}}]{Yang:2021hrl}%
  \BibitemOpen
  \bibfield  {author} {\bibinfo {author} {\bibfnamefont {P.}~\bibnamefont
  {Yang}}, \bibinfo {author} {\bibfnamefont {Y.}~\bibnamefont {Jiang}},
  \bibinfo {author} {\bibfnamefont {S.}~\bibnamefont {Komatsu}},\ and\ \bibinfo
  {author} {\bibfnamefont {J.-B.}\ \bibnamefont {Wu}},\ }\href
  {https://doi.org/10.1007/JHEP01(2022)002} {\bibfield  {journal} {\bibinfo
  {journal} {JHEP}\ }\textbf {\bibinfo {volume} {01}},\ \bibinfo {pages}
  {002}},\ \Eprint {https://arxiv.org/abs/2103.15840} {arXiv:2103.15840
  [hep-th]} \BibitemShut {NoStop}%
\bibitem [{\citenamefont {Kristjansen}\ \emph {et~al.}(2022)\citenamefont
  {Kristjansen}, \citenamefont {Vu},\ and\ \citenamefont
  {Zarembo}}]{Kristjansen:2021abc}%
  \BibitemOpen
  \bibfield  {author} {\bibinfo {author} {\bibfnamefont {C.}~\bibnamefont
  {Kristjansen}}, \bibinfo {author} {\bibfnamefont {D.-L.}\ \bibnamefont
  {Vu}},\ and\ \bibinfo {author} {\bibfnamefont {K.}~\bibnamefont {Zarembo}},\
  }\href {https://doi.org/10.1007/JHEP02(2022)070} {\bibfield  {journal}
  {\bibinfo  {journal} {JHEP}\ }\textbf {\bibinfo {volume} {02}},\ \bibinfo
  {pages} {070}},\ \Eprint {https://arxiv.org/abs/2112.10438} {arXiv:2112.10438
  [hep-th]} \BibitemShut {NoStop}%
\bibitem [{\citenamefont {Gromov}\ and\ \citenamefont
  {Kazakov}(2012)}]{Gromov:2010kf}%
  \BibitemOpen
  \bibfield  {author} {\bibinfo {author} {\bibfnamefont {N.}~\bibnamefont
  {Gromov}}\ and\ \bibinfo {author} {\bibfnamefont {V.}~\bibnamefont
  {Kazakov}},\ }\href {https://doi.org/10.1007/s11005-011-0513-x} {\bibfield
  {journal} {\bibinfo  {journal} {Lett. Math. Phys.}\ }\textbf {\bibinfo
  {volume} {99}},\ \bibinfo {pages} {321} (\bibinfo {year} {2012})},\ \Eprint
  {https://arxiv.org/abs/1012.3996} {arXiv:1012.3996 [hep-th]} \BibitemShut
  {NoStop}%
\bibitem [{\citenamefont {Gromov}\ \emph {et~al.}(2014)\citenamefont {Gromov},
  \citenamefont {Kazakov}, \citenamefont {Leurent},\ and\ \citenamefont
  {Volin}}]{Gromov:2013pga}%
  \BibitemOpen
  \bibfield  {author} {\bibinfo {author} {\bibfnamefont {N.}~\bibnamefont
  {Gromov}}, \bibinfo {author} {\bibfnamefont {V.}~\bibnamefont {Kazakov}},
  \bibinfo {author} {\bibfnamefont {S.}~\bibnamefont {Leurent}},\ and\ \bibinfo
  {author} {\bibfnamefont {D.}~\bibnamefont {Volin}},\ }\href
  {https://doi.org/10.1103/PhysRevLett.112.011602} {\bibfield  {journal}
  {\bibinfo  {journal} {Phys. Rev. Lett.}\ }\textbf {\bibinfo {volume} {112}},\
  \bibinfo {pages} {011602} (\bibinfo {year} {2014})},\ \Eprint
  {https://arxiv.org/abs/1305.1939} {arXiv:1305.1939 [hep-th]} \BibitemShut
  {NoStop}%
\bibitem [{\citenamefont {Cavagli\`a}\ \emph {et~al.}(2019)\citenamefont
  {Cavagli\`a}, \citenamefont {Gromov},\ and\ \citenamefont
  {Levkovich-Maslyuk}}]{Cavaglia:2019pow}%
  \BibitemOpen
  \bibfield  {author} {\bibinfo {author} {\bibfnamefont {A.}~\bibnamefont
  {Cavagli\`a}}, \bibinfo {author} {\bibfnamefont {N.}~\bibnamefont {Gromov}},\
  and\ \bibinfo {author} {\bibfnamefont {F.}~\bibnamefont
  {Levkovich-Maslyuk}},\ }\href {https://doi.org/10.1007/JHEP09(2019)052}
  {\bibfield  {journal} {\bibinfo  {journal} {JHEP}\ }\textbf {\bibinfo
  {volume} {09}},\ \bibinfo {pages} {052}},\ \Eprint
  {https://arxiv.org/abs/1907.03788} {arXiv:1907.03788 [hep-th]} \BibitemShut
  {NoStop}%
\bibitem [{\citenamefont {Ryan}\ and\ \citenamefont
  {Volin}(2019)}]{Ryan:2018fyo}%
  \BibitemOpen
  \bibfield  {author} {\bibinfo {author} {\bibfnamefont {P.}~\bibnamefont
  {Ryan}}\ and\ \bibinfo {author} {\bibfnamefont {D.}~\bibnamefont {Volin}},\
  }\href {https://doi.org/10.1063/1.5085387} {\bibfield  {journal} {\bibinfo
  {journal} {J. Math. Phys.}\ }\textbf {\bibinfo {volume} {60}},\ \bibinfo
  {pages} {032701} (\bibinfo {year} {2019})},\ \Eprint
  {https://arxiv.org/abs/1810.10996} {arXiv:1810.10996 [math-ph]} \BibitemShut
  {NoStop}%
\bibitem [{\citenamefont {Maillet}\ and\ \citenamefont
  {Niccoli}(2018)}]{Maillet:2018bim}%
  \BibitemOpen
  \bibfield  {author} {\bibinfo {author} {\bibfnamefont {J.~M.}\ \bibnamefont
  {Maillet}}\ and\ \bibinfo {author} {\bibfnamefont {G.}~\bibnamefont
  {Niccoli}},\ }\href {https://doi.org/10.1063/1.5050989} {\bibfield  {journal}
  {\bibinfo  {journal} {J. Math. Phys.}\ }\textbf {\bibinfo {volume} {59}},\
  \bibinfo {pages} {091417} (\bibinfo {year} {2018})},\ \Eprint
  {https://arxiv.org/abs/1807.11572} {arXiv:1807.11572 [math-ph]} \BibitemShut
  {NoStop}%
\bibitem [{\citenamefont {Beisert}\ and\ \citenamefont
  {Erkal}(2008)}]{Beisert:2007jv}%
  \BibitemOpen
  \bibfield  {author} {\bibinfo {author} {\bibfnamefont {N.}~\bibnamefont
  {Beisert}}\ and\ \bibinfo {author} {\bibfnamefont {D.}~\bibnamefont
  {Erkal}},\ }\href {https://doi.org/10.1088/1742-5468/2008/03/P03001}
  {\bibfield  {journal} {\bibinfo  {journal} {J. Stat. Mech.}\ }\textbf
  {\bibinfo {volume} {0803}},\ \bibinfo {pages} {P03001} (\bibinfo {year}
  {2008})},\ \Eprint {https://arxiv.org/abs/0711.4813} {arXiv:0711.4813
  [hep-th]} \BibitemShut {NoStop}%
\bibitem [{\citenamefont {Pozsgay}\ \emph {et~al.}(2020)\citenamefont
  {Pozsgay}, \citenamefont {Jiang},\ and\ \citenamefont
  {Tak\'acs}}]{Pozsgay:2019ekd}%
  \BibitemOpen
  \bibfield  {author} {\bibinfo {author} {\bibfnamefont {B.}~\bibnamefont
  {Pozsgay}}, \bibinfo {author} {\bibfnamefont {Y.}~\bibnamefont {Jiang}},\
  and\ \bibinfo {author} {\bibfnamefont {G.}~\bibnamefont {Tak\'acs}},\ }\href
  {https://doi.org/10.1007/JHEP03(2020)092} {\bibfield  {journal} {\bibinfo
  {journal} {JHEP}\ }\textbf {\bibinfo {volume} {03}},\ \bibinfo {pages}
  {092}},\ \Eprint {https://arxiv.org/abs/1911.11118} {arXiv:1911.11118
  [hep-th]} \BibitemShut {NoStop}%
\bibitem [{\citenamefont {de~Leeuw}\ and\ \citenamefont
  {Retore}(2022)}]{deLeeuw:2022}%
  \BibitemOpen
  \bibfield  {author} {\bibinfo {author} {\bibfnamefont {M.}~\bibnamefont
  {de~Leeuw}}\ and\ \bibinfo {author} {\bibfnamefont {A.~L.}\ \bibnamefont
  {Retore}},\ }\href@noop {} {\  (\bibinfo {year} {2022})},\ \Eprint
  {https://arxiv.org/abs/2206.08390} {arXiv:2206.08390 [hep-th]} \BibitemShut
  {NoStop}%
\bibitem [{\citenamefont {Gombor}\ and\ \citenamefont
  {Pozsgay}(2022)}]{Gombor:2022ldb}%
  \BibitemOpen
  \bibfield  {author} {\bibinfo {author} {\bibfnamefont {T.}~\bibnamefont
  {Gombor}}\ and\ \bibinfo {author} {\bibfnamefont {B.}~\bibnamefont
  {Pozsgay}},\ }\href@noop {} {\  (\bibinfo {year} {2022})},\ \Eprint
  {https://arxiv.org/abs/2205.02038} {arXiv:2205.02038 [nlin.SI]} \BibitemShut
  {NoStop}%
\end{thebibliography}%

\pagebreak

\widetext

\newpage
\begin{center}
\textbf{\large Supplemental Materials: Wrapping corrections for long range spin chains}
\end{center}
\setcounter{equation}{0}
\setcounter{figure}{0}
\setcounter{table}{0}
\setcounter{page}{1}
\setcounter{section}{0}
\makeatletter
\renewcommand{\theequation}{S-\arabic{equation}}
\renewcommand{\thefigure}{S-\arabic{figure}}
\renewcommand{\thesection}{S-\Roman{section}}

\section{Review for the integrability of medium range spin chains}

\begin{figure}
\begin{centering}
\includegraphics[width=0.3\textwidth]{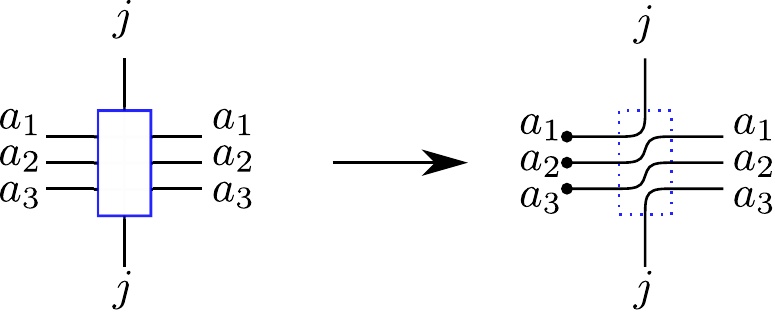}
\par\end{centering}
\caption{Graphical illustration for the unchecked Lax-operator (\ref{eq:unchL})
for $\ell=2$. The right graph shows the regularity property (\ref{eq:reg}).}

\label{fig:unchL}
\end{figure}

In this section we review the algebraic construction of integrable
medium range spin chains \citet{Gombor:2021nhn}. Some readers might
be more familiar to the ''unchecked'' notations for Lax- and $R$-operators
therefore we start with this convention. Assuming that the interaction
range is $\ell+2$ we have to introduce $\ell+1$ auxiliary spaces
labeled by $a_{1},a_{2},\dots,a_{\ell+1}$ and the tensor product
of all of them is labeled by $A=(a_{1},a_{2},\dots,a_{\ell+1})$.
The Lax operator acts on the auxiliary space $A$ and one site of
the quantum space $j$ (see figure \ref{fig:unchL}):
\begin{equation}
L_{A,j}^{(\ell)}(u)\equiv L_{(a_{1},a_{2},\dots,a_{\ell+1}),j}^{(\ell)}(u).\label{eq:unchL}
\end{equation}

We also define the $R$-matrix $R_{A,B}^{(\ell)}(u,v)$ for which
the usual $RLL$-relation is satisfied
\begin{equation}
R_{A,B}^{(\ell)}(u,v)L_{A,j}^{(\ell)}(u)L_{B,j}^{(\ell)}(v)=L_{B,j}^{(\ell)}(v)L_{A,j}^{(\ell)}(u)R_{A,B}^{(\ell)}(u,v),\label{eq:RLLunch}
\end{equation}
where the $A,B$ are the auxiliary space and $j$ is one site of the
quantum space. The associativity of the algebra of $L^{(\ell)}(u)$-s
requires the Yang-Baxter equation
\begin{equation}
R_{A,B}^{(\ell)}(u,v)R_{A,C}^{(\ell)}(u,w)R_{B,C}^{(\ell)}(v,w)=R_{B,C}^{(\ell)}(v,w)R_{A,C}^{(\ell)}(u,w)R_{A,B}^{(\ell)}(u,v).
\end{equation}

\begin{figure}
\begin{centering}
\includegraphics[width=0.55\textwidth]{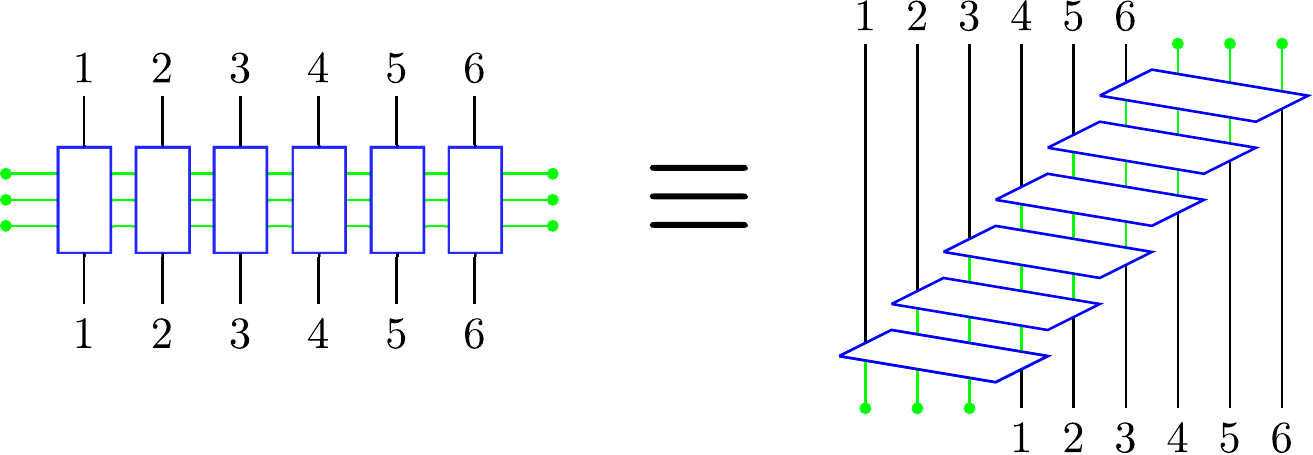}
\par\end{centering}
\caption{Graphical illustration for the unchecked transfer matrix (\ref{eq:unchTransfer})
for $\ell=2$ and $J=6$. The left graph is the most common illustration
of the transfer matrix and the right one shows the same graph but
the placement of the boxes are similar to the graph of the checked
transfer matrix. We can see that the checked transfer matrix contains
extra shifts of the incoming legs.}

\label{fig:unredraw}
\end{figure}

\begin{figure}
\begin{centering}
\includegraphics[width=0.6\textwidth]{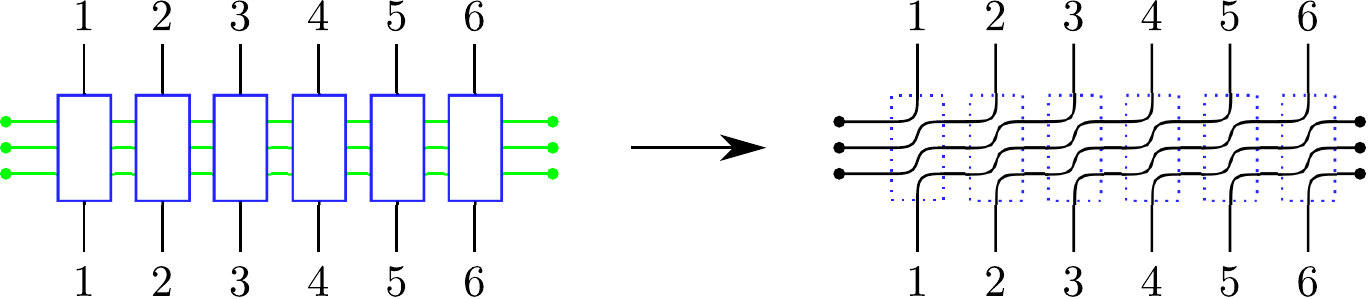}
\par\end{centering}
\caption{Graphical illustration for the $u=0$ limit of the ''unchecked'' transfer
matrix (\ref{eq:tu0}) for $\ell=2$ and $J=6$.}

\label{fig:unchtrans}
\end{figure}

We can also define the transfer matrix
\begin{equation}
T^{(\ell)}(u)=\mathrm{Tr}_{A}\left(L_{A,J}^{(\ell)}(u)L_{A,J-1}^{(\ell)}(u)\dots L_{A,1}^{(\ell)}(u)\right).\label{eq:unchTransfer}
\end{equation}
The commutativity of the transfer matrix 
\begin{equation}
[T^{(\ell)}(u),T^{(\ell)}(v)]=0\label{eq:comT}
\end{equation}
is a simple consequence of the $RLL$-relation. To obtain local conserved
charges we introduce the regularity condition
\begin{equation}
L_{A,j}^{(\ell)}(u)\equiv L_{(a_{1},a_{2},\dots,a_{\ell+1}),j}^{(\ell)}(u)=P_{a_{1},j}P_{a_{2},j}\dots P_{a_{\ell+1},j}\left(1+uh_{a_{1},a_{2},\dots,a_{\ell+1},j}^{(\ell)}+\mathcal{O}(u^{2})\right).\label{eq:reg}
\end{equation}
The first consequence of the regularity condition is that the transfer
matrix is a $\ell+1$ site shift operator (see figure \ref{fig:unchtrans})
\begin{equation}
T^{(\ell)}(0)=U^{\ell+1},\label{eq:tu0}
\end{equation}
where $U$ is the one site shift operator
\begin{equation}
U=P_{1,2}P_{2,3}\dots P_{J-1,J}.
\end{equation}

The second consequence of the regularity condition is that the transfer
matrix gives local Hamiltonian with interaction range $\ell+2$ (assuming
$J\geq\ell+2$):
\begin{equation}
\frac{d}{du}\log T^{(\ell)}(u)\Biggr|_{u=0}=\sum_{j=1}^{J}h_{j}^{(\ell)}=H^{(\ell)}.
\end{equation}

\begin{figure}
\begin{centering}
\includegraphics[width=0.65\textwidth]{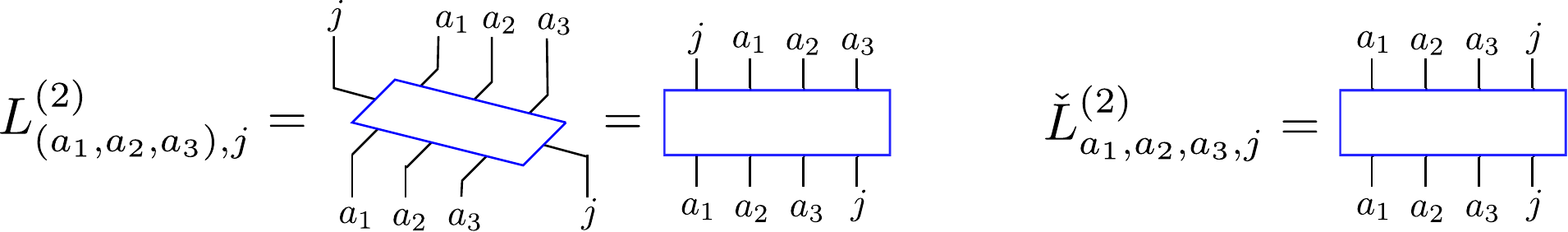}
\par\end{centering}
\caption{Graphical illustration for the ''unchecked'' and ''checked'' Lax-operators
(\ref{eq:chuch}) for $\ell=2$.}

\label{fig:unchVSch}
\end{figure}
We can see that series expansion is more simpler if we introduce the
''checked'' operators as (see figure \ref{fig:unchVSch})
\begin{align}
L_{(a_{1},a_{2},\dots,a_{\ell+1}),j}^{(\ell)}(u) & =P_{a_{1},j}P_{a_{2},j}\dots P_{a_{\ell+1},j}\check{L}_{a_{1},a_{2},\dots,a_{\ell+1},j}^{(\ell)}(u),\label{eq:chuch}\\
R_{(a_{1},\dots,a_{\ell+1}),(b_{1},\dots,b_{\ell+1})}^{(\ell)}(u,v) & =P_{a_{1},b_{1}}\dots P_{a_{\ell+1},b_{\ell+1}}\check{R}_{a_{1},\dots,a_{\ell+1},b_{1},\dots,b_{\ell+1}}^{(\ell)}(u,v).
\end{align}
Substituting back to (\ref{eq:RLLunch}) we obtain the $RLL$-relation
in the ''checked'' convention
\begin{multline}
\check{R}_{a_{2},\dots,a_{\ell+1},b_{1},\dots,b_{\ell+1},j}^{(\ell)}(u,v)\check{L}_{a_{1},a_{2},\dots,a_{\ell+1},b_{1}}^{(\ell)}(u)\check{L}_{b_{1},b_{2},\dots,b_{\ell+1},j}^{(\ell)}(v)=\\
\check{L}_{a_{1},a_{2},\dots,a_{\ell+1},b_{1}}^{(\ell)}(v)\check{L}_{b_{1},b_{2},\dots,b_{\ell+1},j}^{(\ell)}(u)\check{R}_{a_{1},\dots,a_{\ell+1},b_{1},\dots,b_{\ell+1}}^{(\ell)}(u,v).
\end{multline}
Introducing new labeling for the sites as $a_{1},\dots,a_{\ell+1}\leftrightarrow1,\dots,\ell+1$
and $b_{1},\dots,b_{\ell+1}\leftrightarrow\ell+2,\dots,2\ell+2$ and
$j\leftrightarrow2\ell+3$ we obtain that
\begin{equation}
\check{R}_{2,\dots,2\ell+3}^{(\ell)}(u,v)\check{L}_{1,\dots,\ell+2}^{(\ell)}(u)\check{L}_{\ell+2,\dots,2\ell+3}^{(\ell)}(v)=\check{L}_{1,\dots,\ell+2}^{(\ell)}(v)\check{L}_{\ell+2,\dots,2\ell+3}^{(\ell)}(u)\check{R}_{1,\dots,2\ell+2}^{(\ell)}(u,v).
\end{equation}
We can see that the ''checked'' operators always act on neighboring
sites therefore, it is unnecessary to write out all the sites where
the ''checked'' operators act, it is sufficient to label only the
one on the left as
\begin{align}
\check{L}_{j}^{(\ell)}(u) & \equiv\check{L}_{j,\dots,j+\ell+1}^{(\ell)}(u),\\
\check{R}_{j}^{(\ell)}(u,v) & \equiv\check{R}_{j,\dots,j+2\ell+1}^{(\ell)}(u,v).
\end{align}
We can also introduce the ''checked'' version of the transfer matrix
\begin{equation}
\check{T}^{(\ell)}(u)=\widehat{\mathrm{Tr}}_{J,\ell+1}\left(\check{L}_{J}^{(\ell)}(u)\dots\check{L}_{1}^{(\ell)}(u)\right).
\end{equation}
Since the checked Lax-operator has the regularity property $\check{L}_{j}^{(\ell)}(0)=1$
the checked transfer matrix is identity at $u=0$, i.e.
\begin{equation}
\check{T}^{(\ell)}(0)=1.
\end{equation}
We can also show the following connection between the two conventions
for the transfer matrices (see figure \ref{fig:unredraw})
\begin{equation}
T^{(\ell)}(u)=\check{T}^{(\ell)}(u)U^{\ell+1}.\label{eq:connT}
\end{equation}

The commutativity of the unchecked transfer matrix (\ref{eq:comT})
can be derived in the usual way:
\begin{multline}
T^{(\ell)}(u)T^{(\ell)}(v)=\mathrm{Tr}_{A,B}\left(L_{A,J}^{(\ell)}(u)L_{B,J}^{(\ell)}(v)\dots L_{A,1}^{(\ell)}(u)L_{B,1}^{(\ell)}(v)\right)=\\
\mathrm{Tr}_{A,B}\left(R_{A,B}^{(\ell)}(u,v)^{-1}R_{A,B}^{(\ell)}(u,v)L_{A,J}^{(\ell)}(u)L_{B,J}^{(\ell)}(v)\dots L_{A,1}^{(\ell)}(u)L_{B,1}^{(\ell)}(v)\right)=\\
\mathrm{Tr}_{A,B}\left(R_{A,B}^{(\ell)}(u,v)^{-1}L_{B,J}^{(\ell)}(v)L_{A,J}^{(\ell)}(u)\dots L_{B,1}^{(\ell)}(v)L_{A,1}^{(\ell)}(u)R_{A,B}^{(\ell)}(u,v)\right)=\\
\mathrm{Tr}_{A,B}\left(L_{B,J}^{(\ell)}(v)L_{A,J}^{(\ell)}(u)\dots L_{B,1}^{(\ell)}(v)L_{A,1}^{(\ell)}(u)\right)=T^{(\ell)}(v)T^{(\ell)}(u),\label{eq:deriv-1}
\end{multline}
where we used the $RLL$-relation in the third line. From (\ref{eq:connT})
it is clear that the unchecked transfer matrices are also commuting.
We note that the proof can be also done using the checked notations
\citep{Gombor:2022ldb}. It is clear that the derivation (\ref{eq:deriv-1})
is independent of the length of the spin chain $J$ therefore the
transfer matrix is well defined and gives commuting charges even when
$J<\ell+2$.

\section{Properties of the truncated $RLL$-relation}

In this section we investigate the truncated $RLL$-relation
\begin{equation}
\check{\mathcal{R}}_{2}^{(\ell)}\check{\mathcal{L}}_{1}^{(\ell)}(u)\check{\mathcal{L}}_{\ell+2}^{(\ell)}(v)=\check{\mathcal{L}}_{1}^{(\ell)}(v)\check{\mathcal{L}}_{\ell+2}^{(\ell)}(u)\check{\mathcal{R}}_{1}^{(\ell)}+\mathcal{O}(\lambda^{\ell+1}).
\end{equation}
This relation is equivalent with $\ell+1$ different equations
\begin{equation}
\sum_{k=0}^{r}\sum_{l=0}^{r-k}\check{R}_{2}^{(\ell,r-k-l)}\check{L}_{1}^{(k)}(u)\check{L}_{\ell+2}^{(l)}(v)=\sum_{k=0}^{r}\sum_{l=0}^{r-k}\check{L}_{1}^{(k)}(v)\check{L}_{\ell+2}^{(l)}(u)\check{R}_{1}^{(\ell,r-k-l)},\label{eq:rlls}
\end{equation}
for $r=0,1,\dots,\ell$. We can see that the operators $\check{R}_{1}^{(\ell,\ell)},\check{L}_{1}^{(\ell)}(u)$
appear only in the equation $r=\ell$. In the following we will show
that the remaining equations $r=0,1,\dots,\ell-1$ are equivalent
to the equations coming from the previous order
\begin{equation}
\check{\mathcal{R}}_{2}^{(\ell-1)}\check{\mathcal{L}}_{1}^{(\ell-1)}(u)\check{\mathcal{L}}_{\ell+1}^{(\ell-1)}(v)=\check{\mathcal{L}}_{1}^{(\ell-1)}(v)\check{\mathcal{L}}_{\ell+1}^{(\ell-1)}(u)\check{\mathcal{R}}_{1}^{(\ell-1)}+\mathcal{O}(\lambda^{\ell}).\label{eq:Rll}
\end{equation}

At first, let us define the operator
\begin{equation}
\mathcal{X}_{1}^{(\ell)}=\check{\mathcal{L}}_{\ell+1}^{(\ell-1)}(u)\check{\mathcal{R}}_{1}^{(\ell-1)}\check{\mathcal{J}}_{\ell+1}^{(\ell-1)}(v)+\mathcal{O}(\lambda^{\ell}),\label{eq:Xdef}
\end{equation}
with range $2\ell+1$. In the following, we prove that, it satisfies
the $RLL$-relations (\ref{eq:rlls}) for $r=0,\dots,\ell-1$, i.e.,
\begin{equation}
\mathcal{X}_{2}^{(\ell)}\check{\mathcal{L}}_{1}^{(\ell)}(u)\check{\mathcal{L}}_{\ell+2}^{(\ell)}(v)=\check{\mathcal{L}}_{1}^{(\ell)}(v)\check{\mathcal{L}}_{\ell+2}^{(\ell)}(u)\mathcal{X}_{1}^{(\ell)}+\mathcal{O}(\lambda^{\ell}).\label{eq:XLL}
\end{equation}
Since $\check{\mathcal{L}}_{1}^{(\ell)}(u)=\check{\mathcal{L}}_{1}^{(\ell-1)}(u)+\mathcal{O}(\lambda^{\ell})$,
the equation (\ref{eq:XLL}) is equivalent to 
\begin{equation}
\mathcal{X}_{2}^{(\ell)}\check{\mathcal{L}}_{1}^{(\ell-1)}(u)\check{\mathcal{L}}_{\ell+2}^{(\ell-1)}(v)=\check{\mathcal{L}}_{1}^{(\ell-1)}(v)\check{\mathcal{L}}_{\ell+2}^{(\ell-1)}(u)\mathcal{X}_{1}^{(\ell)}+\mathcal{O}(\lambda^{\ell}).\label{eq:XLL2}
\end{equation}
Let us substitute (\ref{eq:Xdef}) to (\ref{eq:XLL2}).
\begin{equation}
\check{\mathcal{L}}_{\ell+2}^{(\ell-1)}(u)\check{\mathcal{R}}_{2}^{(\ell-1)}\check{\mathcal{J}}_{\ell+2}^{(\ell-1)}(v)\check{\mathcal{L}}_{1}^{(\ell-1)}(u)\check{\mathcal{L}}_{\ell+2}^{(\ell-1)}(v)=\check{\mathcal{L}}_{1}^{(\ell-1)}(v)\check{\mathcal{L}}_{\ell+2}^{(\ell-1)}(u)\check{\mathcal{L}}_{\ell+1}^{(\ell-1)}(u)\check{\mathcal{R}}_{1}^{(\ell-1)}\check{\mathcal{J}}_{\ell+1}^{(\ell-1)}(v)+\mathcal{O}(\lambda^{\ell}).
\end{equation}
Since $\check{\mathcal{L}}_{1}^{(\ell-1)}(u)=\check{\mathcal{L}}_{1,2,\dots,\ell+1}^{(\ell-1)}(u)$,
we have commuting operators
\begin{equation}
\left[\check{\mathcal{L}}_{1}^{(\ell-1)}(u),\check{\mathcal{L}}_{\ell+2}^{(\ell-1)}(v)\right]=0,
\end{equation}
therefore we can obtain that
\begin{equation}
\check{\mathcal{R}}_{2}^{(\ell-1)}\check{\mathcal{L}}_{1}^{(\ell-1)}(u)=\check{\mathcal{L}}_{1}^{(\ell-1)}(v)\check{\mathcal{L}}_{\ell+1}^{(\ell-1)}(u)\check{\mathcal{R}}_{1}^{(\ell-1)}\check{\mathcal{J}}_{\ell+1}^{(\ell-1)}(v)+\mathcal{O}(\lambda^{\ell}),
\end{equation}
where we used that the $\check{\mathcal{J}}$ is the inverse of $\check{\mathcal{L}}$
i.e.
\begin{equation}
\check{\mathcal{J}}_{1}^{(\ell-1)}(u)\check{\mathcal{L}}_{1}^{(\ell-1)}(u)=1+\mathcal{O}(\lambda^{\ell}).\label{eq:inv}
\end{equation}
Multiplying $\check{\mathcal{L}}_{\ell+1}^{(\ell-1)}(v)$ from the
right we obtain that
\begin{equation}
\check{\mathcal{R}}_{2}^{(\ell-1)}\check{\mathcal{L}}_{1}^{(\ell-1)}(u)\check{\mathcal{L}}_{\ell+1}^{(\ell-1)}(v)=\check{\mathcal{L}}_{1}^{(\ell-1)}(v)\check{\mathcal{L}}_{\ell+1}^{(\ell-1)}(u)\check{\mathcal{R}}_{1}^{(\ell-1)}+\mathcal{O}(\lambda^{\ell}),
\end{equation}
which is satisfied by the initial condition (\ref{eq:Rll}) therefore
we just proved (\ref{eq:XLL}). Since (\ref{eq:XLL}) is the defining
equation of $\check{\mathcal{R}}^{(\ell)}$ up to order $\mathcal{O}(\lambda^{\ell})$
we can identify $\check{\mathcal{R}}^{(\ell)}$ with $\mathcal{X}^{(\ell)}$
up to this order i.e.
\begin{equation}
\check{\mathcal{R}}_{1}^{(\ell)}=\mathcal{X}_{1}^{(\ell)}+\mathcal{O}(\lambda^{\ell})=\check{\mathcal{L}}_{\ell+1}^{(\ell-1)}(u)\check{\mathcal{R}}_{1}^{(\ell-1)}\check{\mathcal{J}}_{\ell+1}^{(\ell-1)}(v)+\mathcal{O}(\lambda^{\ell}).
\end{equation}

Let us summarize what we obtained. If we have operators 
\begin{equation}
\check{\mathcal{R}}_{1}^{(\ell-1)}=\sum_{k=0}^{\ell-1}\lambda^{k}\check{R}_{1}^{(\ell-1,k)},\qquad\check{\mathcal{L}}_{1}^{(\ell-1)}=\sum_{k=0}^{\ell-1}\lambda^{k}\check{L}_{1}^{(k)},\qquad\check{\mathcal{J}}_{1}^{(\ell-1)}=\sum_{k=0}^{\ell-1}\lambda^{k}\check{J}_{1}^{(k)},
\end{equation}
which satisfy the truncated $RLL$-equations (\ref{eq:Rll}) and inversion
relation (\ref{eq:inv}) then the operators 
\begin{equation}
\check{R}_{1}^{(\ell,r)}=\sum_{k=0}^{r}\sum_{l=0}^{r-k}\check{L}_{\ell+1}^{(k)}(u)\check{R}_{1}^{(\ell-1,r-k-l)}\check{\mathcal{J}}_{\ell+1}^{(l)}(v),\label{eq:Req}
\end{equation}
for $r=0,\dots,\ell-1$ satisfy the next level truncated $RLL$-equations
(\ref{eq:rlls}) for $r=0,\dots,\ell-1$. From the remaining $r=\ell$
equation
\begin{equation}
\sum_{k=0}^{\ell}\sum_{l=0}^{\ell-k}\check{R}_{2}^{(\ell,\ell-k-l)}\check{L}_{1}^{(k)}(u)\check{L}_{\ell+2}^{(l)}(v)=\sum_{k=0}^{\ell}\sum_{l=0}^{\ell-k}\check{L}_{1}^{(k)}(v)\check{L}_{\ell+2}^{(l)}(u)\check{R}_{1}^{(\ell,\ell-k-l)},\label{eq:RLLeq}
\end{equation}
we can obtain the operators $\check{R}_{1}^{(\ell,\ell)}$ and $\check{L}_{1}^{(\ell)}(u)$
which are the only remaining terms of the operators 
\begin{equation}
\check{\mathcal{R}}_{1}^{(\ell)}=\sum_{k=0}^{\ell}\lambda^{k}\check{R}_{1}^{(\ell,k)},\qquad\check{\mathcal{L}}_{1}^{(\ell)}=\sum_{k=0}^{\ell}\lambda^{k}\check{L}_{1}^{(k)}.
\end{equation}
Defining the operator
\begin{equation}
\check{J}_{1}^{(\ell)}(u)=-\left(\sum_{k=0}^{\ell-1}\check{J}_{1}^{(k)}(u)\check{L}_{1}^{(\ell-k)}(u)\right)\check{J}_{1}^{(0)}(u),\label{eq:Jeq}
\end{equation}
we also obtain the inverse of the Lax in the next level
\begin{equation}
\check{\mathcal{J}}_{1}^{(\ell)}=\sum_{k=0}^{\ell}\lambda^{k}\check{J}_{1}^{(k)}.
\end{equation}

\begin{figure}
\begin{centering}
\includegraphics[width=0.2\columnwidth]{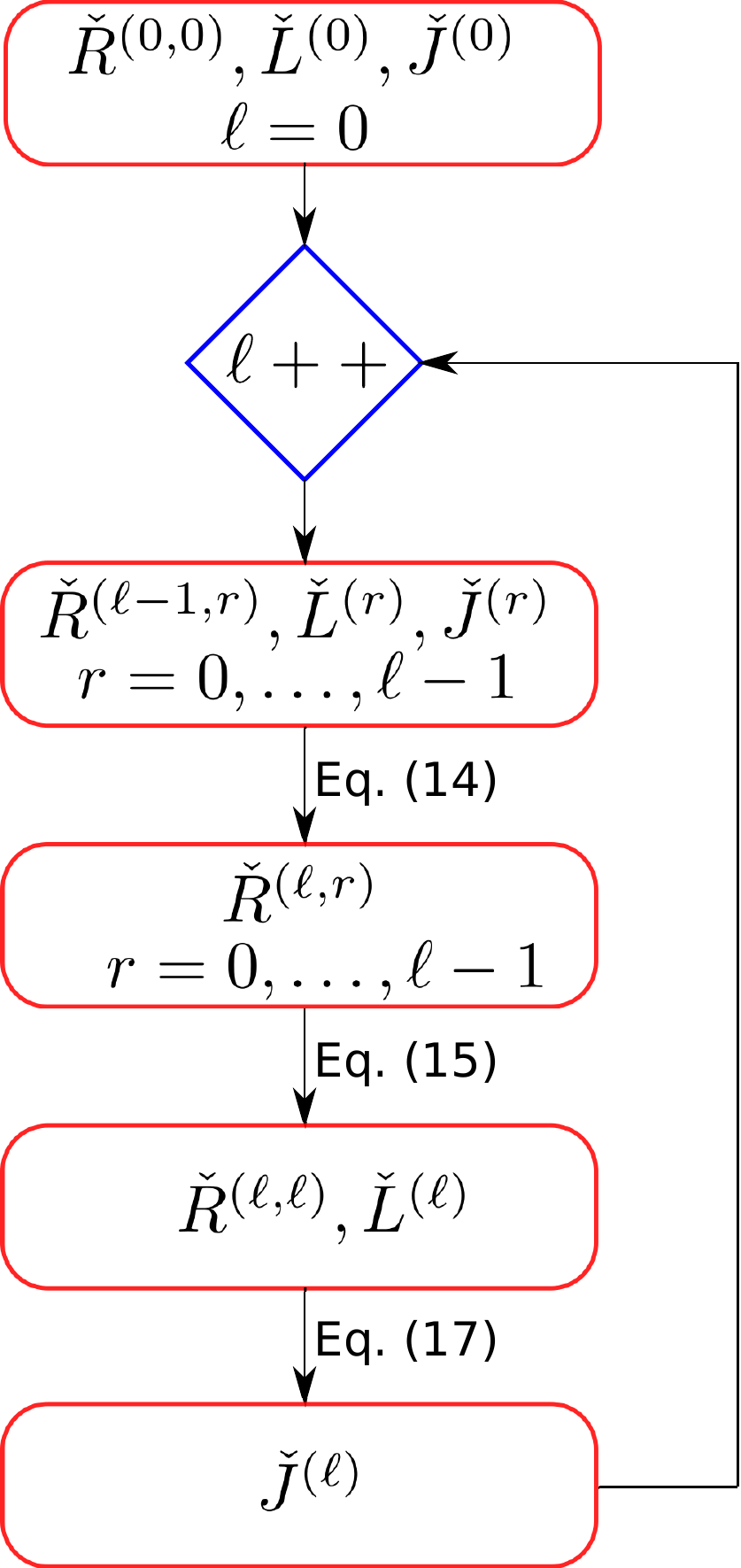}\hspace{0.2\textwidth}\includegraphics[width=0.2\columnwidth]{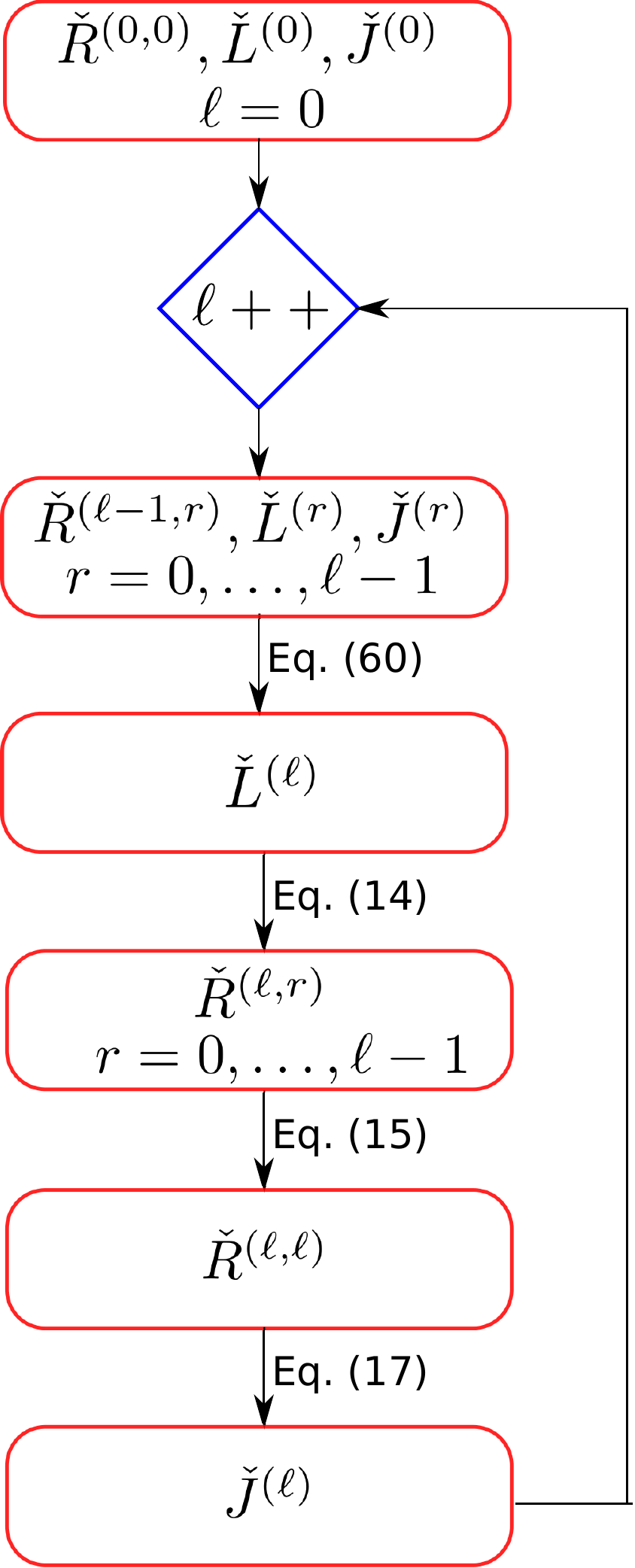}
\par\end{centering}
\caption{Two strategy for finding the Lax-operators and $R$-matrices.}

\centering{}\label{fig:req}
\end{figure}

We can see that we obtained a recursion procedure to solve the truncated
$RLL$-relations (see the left figure of \ref{fig:req} for a summary).

\section{Identities of the twisted trace}

\begin{figure}
\begin{centering}
\includegraphics[width=0.8\columnwidth]{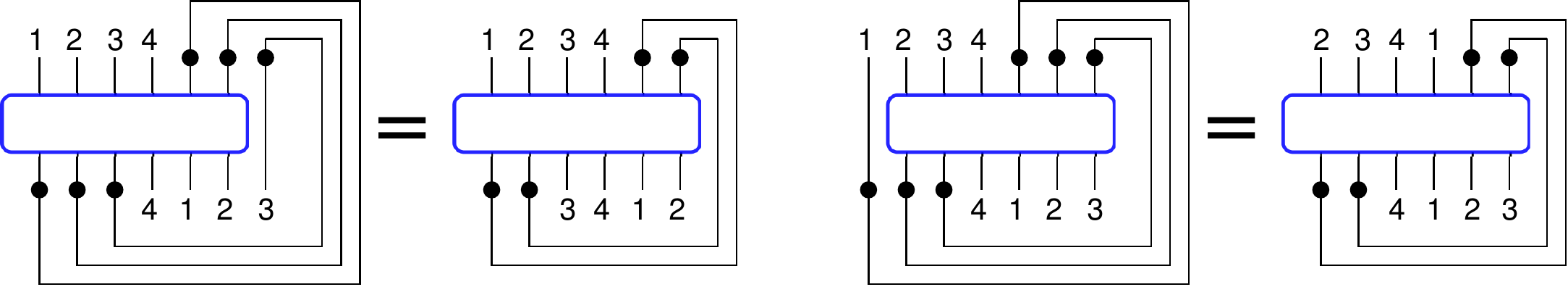}
\par\end{centering}
\caption{The graphical proof of identities (\ref{eq:id1}) and (\ref{eq:id2})
for $J=4$, $\ell=3$.}

\centering{}\label{fig:id1}
\end{figure}
In this section, we show some useful identities of the twisted trace
$\widehat{\mathrm{Tr}}_{J,\ell}$. Let $X_{1}^{(J+\ell-2)}$ be a
range $J+\ell$ operator. We can easily show that (see also figure
\ref{fig:id1})
\begin{equation}
\widehat{\mathrm{Tr}}_{J,\ell+k}\left(X_{1}^{(J+\ell-2)}\right)=\widehat{\mathrm{Tr}}_{J,\ell}\left(X_{1}^{(J+\ell-2)}\right),\label{eq:id1}
\end{equation}
for $k\geq0$. We can obtain analogous identities for the shifted
operators
\begin{equation}
\widehat{\mathrm{Tr}}_{J,\ell+k}\left(X_{l}^{(J+\ell-2)}\right)=\widehat{\mathrm{Tr}}_{J,\ell+l-1}\left(X_{l}^{(J+\ell-2)}\right),
\end{equation}
for $k\geq l-1$. We can see that the action of $\widehat{\mathrm{Tr}}_{J,\ell}$
does not depend on $\ell$ for large enough $\ell$ therefore in the
following we use a shorter notation 
\begin{equation}
\widehat{\mathrm{Tr}}_{J}\left(X_{k}^{(\ell)}\right):=\widehat{\mathrm{Tr}}_{J,n}\left(X_{k}^{(\ell)}\right),
\end{equation}
for $n\geq\max(1,k+\ell+2-J)$.

Another useful identity is (see also figure \ref{fig:id1})
\begin{equation}
\widehat{\mathrm{Tr}}_{J}\left(X_{k}^{(\ell)}\right)=U_{J}^{k-1}\widehat{\mathrm{Tr}}_{J}\left(X_{1}^{(\ell)}\right)U_{J}^{1-k},\label{eq:id2}
\end{equation}
where we used the shift operator
\begin{equation}
U_{J}=P_{1,2}P_{2,3}\dots P_{J-1,J}.
\end{equation}
Using these definitions, the wrapped Hamiltonians have the following
equivalent forms
\begin{equation}
\mathcal{H}^{J}=\sum_{j=1}^{J}\sum_{\ell=0}^{\infty}\tilde{\mathcal{X}}_{j}^{(J,\ell)}=\sum_{j=1}^{J}\sum_{\ell=0}^{\infty}\widehat{\mathrm{Tr}}_{J}\left(\mathcal{X}_{j}^{(\ell)}\right)=\sum_{j=1}^{J}\sum_{\ell=0}^{\infty}U_{J}^{j-1}\widehat{\mathrm{Tr}}_{J}\left(\mathcal{X}_{1}^{(\ell)}\right)U_{J}^{1-j}.\label{eq:finiteHsup}
\end{equation}

Now let us continue with the identities corresponding to the twisted
trace of products of operators:
\[
\widehat{\mathrm{Tr}}_{J}\left(\mathcal{Y}_{k}^{(m)}\mathcal{X}_{1}^{(m)}\right),\qquad\text{or}\qquad\widehat{\mathrm{Tr}}_{J}\left(\mathcal{Y}_{1}^{(m)}\mathcal{X}_{k}^{(m)}\right),
\]
for $k=1,\dots,m+2$ where $\mathcal{X}_{1}^{(m)}$ and $\mathcal{Y}_{1}^{(m)}$
are range $m+2$ operators. In the following we assume that $m+2>J$
but our formulas work for shorter operators as well, since a shorter
operator $\mathcal{X}_{1}^{(n)}\equiv\mathcal{X}_{1,\dots n+1}^{(n)}$
for $n<m$ always can be considered as a range $m+2$ operator $\mathcal{X}_{1}^{(m)}\equiv\mathcal{X}_{1,\dots m+1}^{(m)}$
which acts identically on the sites $n+2,\dots,m+1$ i.e. $\mathcal{X}_{1,\dots m+1}^{(m)}:=\mathcal{X}_{1,\dots n+1}^{(n)}$.

The first class of identities is

\begin{equation}
\widehat{\mathrm{Tr}}_{J}\left(\mathcal{Y}_{k}^{(m)}\mathcal{X}_{1}^{(m)}\right)=\widehat{\mathrm{Tr}}_{J}\left(\mathcal{X}_{1+J}^{(m)}\mathcal{Y}_{k}^{(m)}\right)=U_{J}^{k-1}\widehat{\mathrm{Tr}}_{J}\left(\mathcal{X}_{2+J-k}^{(m)}\mathcal{Y}_{1}^{(m)}\right)U_{J}^{1-k},\label{eq:Id1}
\end{equation}
where $k=1,\dots,J$.

The second class of identities is

\begin{equation}
\widehat{\mathrm{Tr}}_{J}\left(\mathcal{Y}_{k}^{(m)}\mathcal{X}_{1}^{(n)}\right)=\widehat{\mathrm{Tr}}_{J}\left(\mathcal{X}_{1+J}^{(m)}\mathcal{Y}_{k}^{(m)}\right)=\widehat{\mathrm{Tr}}_{J}\left(\mathcal{X}_{1}^{(m)}\mathcal{Y}_{k-J}^{(m)}\right),\label{eq:Id2}
\end{equation}
for $k=J+1,\dots,m+2$.

The third class of identities is
\begin{equation}
\widehat{\mathrm{Tr}}_{J}\left(\mathcal{Y}_{1}^{(m)}\mathcal{X}_{k}^{(m)}\right)=\widehat{\mathrm{Tr}}_{J}\left(\mathcal{X}_{k+J}^{(m)}\mathcal{Y}_{1}^{(m)}\right),\label{eq:Id3}
\end{equation}
for $k=1,\dots,m+2-J$.

The fourth class of identities is
\begin{equation}
\widehat{\mathrm{Tr}}_{J}\left(\mathcal{Y}_{1}^{(m)}\mathcal{X}_{k}^{(m)}\right)=\widehat{\mathrm{Tr}}_{J}\left(\mathcal{Y}_{1}^{(m)}\right)\widehat{\mathrm{Tr}}_{J}\left(\mathcal{X}_{k}^{(m)}\right),\label{eq:Id4}
\end{equation}
for $k=m+3-J,\dots,m+2$.

\begin{figure}
\begin{centering}
\includegraphics[width=1\columnwidth]{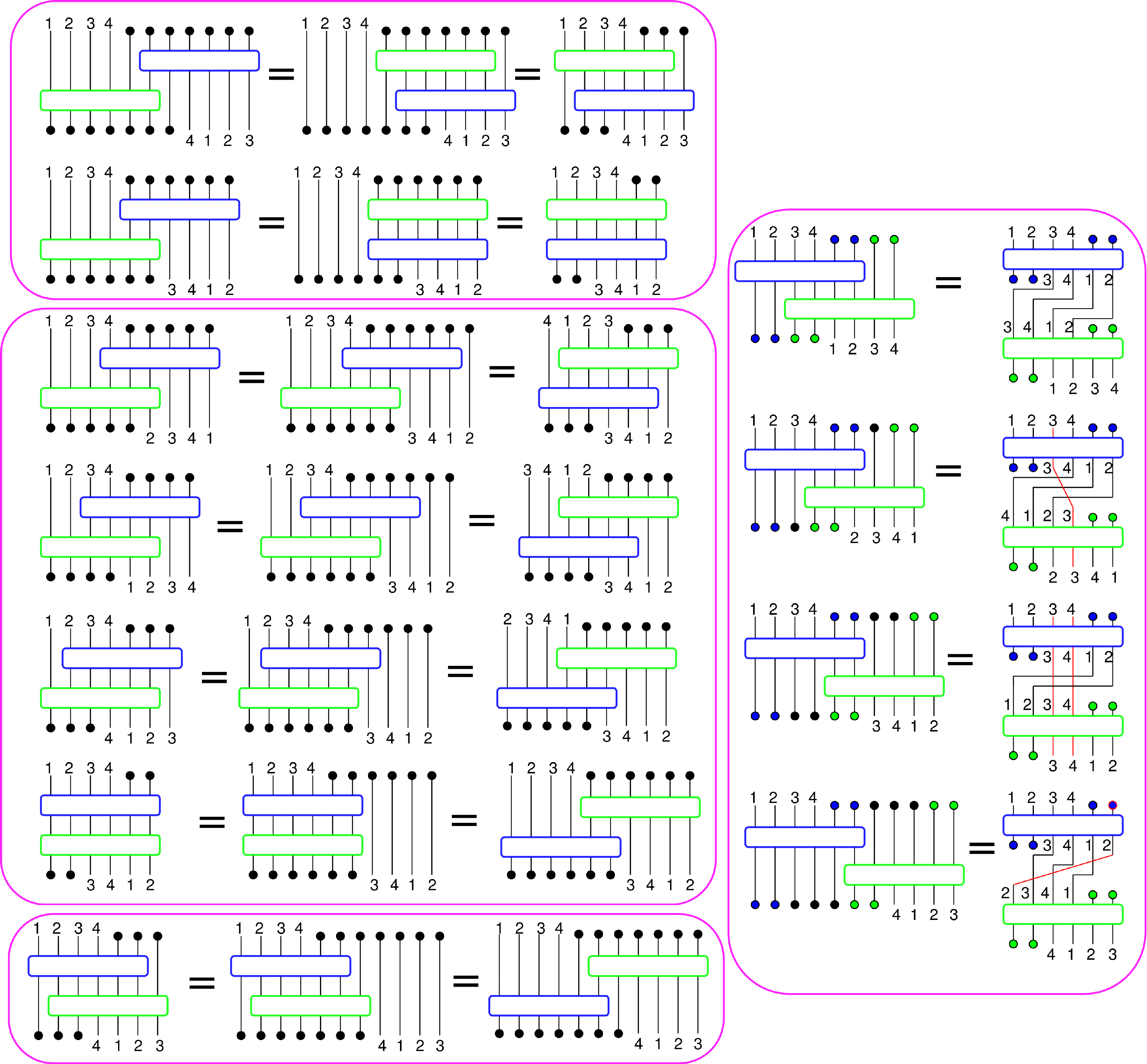}
\par\end{centering}
\caption{The graphical proof of identities (\ref{eq:Id1})-(\ref{eq:Id4})
for $J=4$, $m=4$.}

\centering{}\label{fig:id2}
\end{figure}

The graphical proof of these identities are showed in figure \ref{fig:id2}
for $J=4$, $m=4$.

\section{The $\epsilon$-dependence of the finite volume Hamiltonians}

Let us take an asymptotic Hamiltonian $\mathcal{H}^{0,\infty}$ and
a local operator
\begin{equation}
\mathcal{\mathcal{Y}}=\sum_{j=-\infty}^{\infty}\mathcal{Y}_{j}^{(\ell)},
\end{equation}
where $\mathcal{Y}_{1}^{(\ell)}$ is an $\lambda$ dependent range
$\ell+2$ operator. We can obtain a new integrable asymptotic Hamiltonian
by the rotation
\begin{equation}
\mathcal{H}^{a,\infty}=\exp(a\mathcal{\mathcal{Y}})\mathcal{H}^{0,\infty}\exp(-a\mathcal{\mathcal{Y}})=\sum_{j=-\infty}^{\infty}\sum_{r=0}^{\infty}\mathcal{X}_{j}^{a,(r)}=\sum_{r=0}^{\infty}\mathcal{X}^{a,(r)},
\end{equation}
where $\mathcal{X}_{1}^{a,(r)}$ is a range $r+2$ operator. In this
section we investigate the connection between the wrapped Hamiltonians
$\mathcal{H}^{a,J}$ and $\mathcal{H}^{0,J}$ corresponding to the
asymptotic ones $\mathcal{H}^{a,\infty}$ and $\mathcal{H}^{0,\infty}$,
respectively. The finite volume Hamiltonian reads as
\begin{equation}
\mathcal{H}^{a,J}=\sum_{j=1}^{J}\sum_{\ell=0}^{\infty}\widehat{\mathrm{Tr}}_{J}\left(\mathcal{X}_{j}^{a,(\ell)}\right).\label{eq:finH}
\end{equation}

At first, let us take the derivative of the asymptotic Hamiltonian
with respect to $a$.
\begin{equation}
\frac{d}{da}\mathcal{H}^{a,\infty}=\left[\mathcal{Y},\mathcal{H}^{a,\infty}\right]=\sum_{r=0}^{\infty}\left[\mathcal{Y},\mathcal{X}^{a,(r)}\right].\label{eq:deriv}
\end{equation}
Let us calculate a term from the r.h.s.

\begin{align}
\left[\mathcal{Y},\mathcal{X}^{a,(r)}\right] & =\sum_{j}\left[\mathcal{Y}_{j}^{(\ell)},\mathcal{X}_{j}^{a,(r)}+\mathcal{X}_{j+1}^{a,(r)}+\dots\mathcal{X}_{j+\ell+1}^{a,(r)}+\mathcal{X}_{j-1}^{a,(r)}+\dots+\mathcal{X}_{j-r-1}^{a,(r)}\right]=\nonumber \\
 & =\sum_{j}\left[\mathcal{Y}_{j}^{(\ell)},\mathcal{X}_{j}^{a,(r)}+\mathcal{X}_{j+1}^{a,(r)}+\dots\mathcal{X}_{j+\ell+1}^{a,(r)}\right]+\left[\mathcal{Y}_{j+1}^{(\ell)}+\dots+\mathcal{Y}_{j+r+1}^{(\ell)},\mathcal{X}_{j}^{a,(r)}\right].
\end{align}
Introducing the operator
\begin{equation}
\mathcal{M}_{j}^{(r+\ell+1)}=\left[\mathcal{Y}_{j}^{(\ell)},\mathcal{X}_{j}^{a,(r)}+\mathcal{X}_{j+1}^{a,(r)}+\dots\mathcal{X}_{j+\ell+1}^{a,(r)}\right]+\left[\mathcal{Y}_{j+1}^{(\ell)}+\dots+\mathcal{Y}_{j+r+1}^{(\ell)},\mathcal{X}_{j}^{a,(r)}\right],
\end{equation}
the equation (\ref{eq:deriv}) can be written as
\begin{equation}
\sum_{j=-\infty}^{\infty}\sum_{r=0}^{\infty}\frac{d}{da}\mathcal{X}_{j}^{a,(r)}=\sum_{j=-\infty}^{\infty}\sum_{r=0}^{\infty}\mathcal{M}_{j}^{(\ell+r+1)},
\end{equation}
from which we can obtain that
\begin{equation}
\sum_{r=0}^{\infty}\frac{d}{da}\mathcal{X}_{j}^{a,(r)}=\sum_{r=0}^{\infty}\mathcal{M}_{j}^{(\ell+r+1)}+\sum_{r=0}^{\infty}\left(\mathcal{F}_{j}^{(r)}-\mathcal{F}_{j+1}^{(r)}\right),\label{eq:derivdens}
\end{equation}
where $\mathcal{F}_{1}^{(r)}$ is a range $r+2$ operator. Now let
us calculate the derivative of the wrapped Hamiltonian
\begin{equation}
\frac{d}{da}\mathcal{H}^{a,J}=\sum_{j=1}^{J}\widehat{\mathrm{Tr}}_{J}\left(\sum_{r=0}^{\infty}\frac{d}{da}\mathcal{X}_{j}^{a,(r)}\right)=\sum_{j=1}^{J}\sum_{r=0}^{\infty}\widehat{\mathrm{Tr}}_{J}\left(\mathcal{M}_{j}^{(\ell+r+1)}\right)+\sum_{j=1}^{J}\sum_{r=0}^{\infty}\widehat{\mathrm{Tr}}_{J}\left(\mathcal{F}_{j}^{(r)}-\mathcal{F}_{j+1}^{(r)}\right),
\end{equation}
where we used (\ref{eq:finH}) and (\ref{eq:derivdens}). Since $U_{J}^{J}=1$,
the second sum is vanishing
\begin{equation}
\sum_{j=1}^{J}\sum_{r=0}^{\infty}\widehat{\mathrm{Tr}}_{J}\left(\mathcal{F}_{j}^{(r)}-\mathcal{F}_{j+1}^{(r)}\right)=\sum_{j=1}^{J}U_{J}^{j-1}\sum_{r=0}^{\infty}\widehat{\mathrm{Tr}}_{J}\left(\mathcal{F}_{1}^{(r)}\right)U_{J}^{1-j}-\sum_{j=1}^{J}U_{J}^{j}\sum_{r=0}^{\infty}\widehat{\mathrm{Tr}}_{J}\left(\mathcal{F}_{1}^{(r)}\right)U_{J}^{-j}=0,
\end{equation}
therefore the derivative simplifies as
\begin{equation}
\frac{d}{da}\mathcal{H}^{a,J}=\sum_{r=0}^{\infty}\sum_{j=1}^{J}\tilde{\mathcal{M}}_{j}^{(\ell+r+1)},\label{eq:finiteDeriv}
\end{equation}
where we defined that
\begin{equation}
\tilde{\mathcal{M}}_{j}^{(\ell+r+1)}=\widehat{\mathrm{Tr}}_{J}\left(\mathcal{M}_{j}^{(\ell+r+1)}\right).
\end{equation}
Let us fix the number $r$ and let $m=\max(r,\ell)$. We can also
enlarge the ranges of operators $\mathcal{Y}$ and $\mathcal{X}$
to $m$ as
\begin{equation}
\mathcal{Y}_{1}^{(m)}:=\mathcal{Y}_{1}^{(\ell)},\qquad\mathcal{X}_{1}^{a,(m)}:=\mathcal{X}_{1}^{a,(r)},
\end{equation}
and we can obtain that
\begin{equation}
\mathcal{M}_{1}^{(\ell+r+1)}=\mathcal{M}_{1}^{(2m+1)}:=\left[\mathcal{Y}_{1}^{(m)},\mathcal{X}_{1}^{a,(m)}+\mathcal{X}_{2}^{a,(m)}+\dots+\mathcal{X}_{m+2}^{a,(m)}\right]+\left[\mathcal{Y}_{2}^{(m)}+\dots+\mathcal{Y}_{m+2}^{(m)},\mathcal{X}_{1}^{a,(m)}\right].
\end{equation}
Let us rearrange the operators as
\begin{equation}
\mathcal{M}_{1}^{(2m+1)}=\mathcal{A}^{1}-\mathcal{A}^{2}+\mathcal{B}^{1}-\mathcal{B}^{2}+\mathcal{C}^{1}-\mathcal{C}^{2}+\mathcal{D}^{1}-\mathcal{D}^{2},\label{eq:Mop}
\end{equation}
where
\begin{align}
\mathcal{A}^{1} & =\sum_{k=1}^{J}\mathcal{Y}_{k}^{(m)}\mathcal{X}_{1}^{a,(m)}, & \mathcal{A}^{2} & =\sum_{k=2}^{J+1}\mathcal{X}_{k}^{a,(m)}\mathcal{Y}_{1}^{(m)},\nonumber \\
\mathcal{B}^{1} & =\sum_{k=J+1}^{m+2}\mathcal{Y}_{k}^{(m)}\mathcal{X}_{1}^{a,(m)}, & \mathcal{B}^{2} & =\sum_{k=1}^{m+2-J}\mathcal{X}_{1}^{a,(m)}\mathcal{Y}_{k}^{(m)},\nonumber \\
\mathcal{C}^{1} & =\sum_{k=2}^{m+2-J}\mathcal{Y}_{1}^{(m)}\mathcal{X}_{k}^{a,(m)}, & \mathcal{C}^{2} & =\sum_{k=J+2}^{m+2}\mathcal{X}_{k}^{a,(m)}\mathcal{Y}_{1}^{(m)},\\
\mathcal{D}^{1} & =\sum_{k=m+3-J}^{m+2}\mathcal{Y}_{1}^{(m)}\mathcal{X}_{k}^{a,(m)}, & \mathcal{D}^{2} & =\sum_{k=m+3-J}^{m+2}\mathcal{X}_{1}^{a,(m)}\mathcal{Y}_{k}^{(m)}.\nonumber 
\end{align}
Using the identities (\ref{eq:Id2})-(\ref{eq:Id4}) we can obtain
that
\begin{align}
\widehat{\mathrm{Tr}}_{J}\left(\mathcal{B}^{1}\right) & =\widehat{\mathrm{Tr}}_{J}\left(\mathcal{B}^{2}\right),\\
\widehat{\mathrm{Tr}}_{J}\left(\mathcal{C}^{1}\right) & =\widehat{\mathrm{Tr}}_{J}\left(\mathcal{C}^{2}\right),\\
\widehat{\mathrm{Tr}}_{J}\left(\mathcal{D}^{1}\right) & =\sum_{k=1}^{J}\widehat{\mathrm{Tr}}_{J}\left(\mathcal{Y}_{1}^{(m)}\right)\widehat{\mathrm{Tr}}_{J}\left(\mathcal{X}_{k}^{a,(m)}\right),\\
\widehat{\mathrm{Tr}}_{J}\left(\mathcal{D}^{2}\right) & =\sum_{k=1}^{J}\widehat{\mathrm{Tr}}_{J}\left(\mathcal{X}_{1}^{a,(m)}\right)\widehat{\mathrm{Tr}}_{J}\left(\mathcal{Y}_{k}^{(m)}\right).
\end{align}
Using the identities
\begin{equation}
\widehat{\mathrm{Tr}}_{J}\left(\mathcal{X}_{k}^{a,(m)}\right)=\widehat{\mathrm{Tr}}_{J}\left(\mathcal{X}_{k}^{a,(r)}\right)\qquad\widehat{\mathrm{Tr}}_{J}\left(\mathcal{Y}_{k}^{(m)}\right)=\widehat{\mathrm{Tr}}_{J}\left(\mathcal{Y}_{k}^{(\ell)}\right),
\end{equation}
we can obtain that
\begin{equation}
\widehat{\mathrm{Tr}}_{J}\left(\mathcal{D}^{1}\right)=\sum_{k=1}^{J}\tilde{\mathcal{Y}}_{1}^{(J,\ell)}\tilde{\mathcal{X}}_{k}^{a,(J,r)},\qquad\widehat{\mathrm{Tr}}_{J}\left(\mathcal{D}^{2}\right)=\sum_{k=1}^{J}\tilde{\mathcal{X}}_{1}^{a,(J,r)}\tilde{\mathcal{Y}}_{k}^{(J,\ell)},
\end{equation}
where we defined the wrapped operators
\begin{equation}
\tilde{\mathcal{X}}_{k}^{a,(J,\ell)}=\widehat{\mathrm{Tr}}_{J}\left(\mathcal{X}_{k}^{a,(\ell)}\right),\qquad\tilde{\mathcal{Y}}_{k}^{(J,\ell)}=\widehat{\mathrm{Tr}}_{J}\left(\mathcal{Y}_{k}^{(\ell)}\right).
\end{equation}
Using the identity (\ref{eq:Id1}), we can obtain that
\begin{equation}
\mathcal{A}^{1}=\sum_{k=1}^{J}U_{J}^{k-1}\widehat{\mathrm{Tr}}_{J}\left(\mathcal{X}_{2+J-k}^{a,(m)}\mathcal{Y}_{1}^{(m)}\right)U_{J}^{1-k}=\sum_{k=2}^{J+1}U_{J}^{1-k}\widehat{\mathrm{Tr}}_{J}\left(\mathcal{X}_{k}^{a,(m)}\mathcal{Y}_{1}^{(m)}\right)U_{J}^{k-1}.
\end{equation}
Substituting back to (\ref{eq:Mop}), we obtain that
\begin{equation}
\tilde{\mathcal{M}}_{1}^{(\ell+r+1)}=\sum_{k=1}^{J}\left(\tilde{\mathcal{Y}}_{1}^{(J,\ell)}\tilde{\mathcal{X}}_{k}^{a,(J,r)}-\tilde{\mathcal{X}}_{1}^{a,(J,r)}\tilde{\mathcal{Y}}_{k}^{(J,\ell)}\right)+\sum_{k=2}^{J+1}\left(\mathcal{U}_{J}^{1-k}\widehat{\mathrm{Tr}}_{J}\left(\mathcal{X}_{k}^{a,(m)}\mathcal{Y}_{1}^{(m)}\right)\mathcal{U}_{J}^{k-1}-\widehat{\mathrm{Tr}}_{J}\left(\mathcal{X}_{k}^{a,(m)}\mathcal{Y}_{1}^{(m)}\right)\right),
\end{equation}
therefore whole sum simplifies as
\begin{equation}
\sum_{j=1}^{J}\tilde{\mathcal{M}}_{j}^{(\ell+r+1)}=\sum_{j,k=1}^{J}\left[\tilde{\mathcal{Y}}_{j}^{(J,\ell)},\tilde{\mathcal{X}}_{k}^{a,(J,r)}\right]=\sum_{j=1}^{J}\left[\mathcal{Y}^{J},\tilde{\mathcal{X}}_{j}^{a,(J,r)}\right],
\end{equation}
where we defined that
\begin{equation}
\mathcal{Y}^{J}=\sum_{j=1}^{J}\tilde{\mathcal{Y}}_{j}^{(J,\ell)}.
\end{equation}
Substituting back to (\ref{eq:finiteDeriv}) we just obtained a differential
equation for the finite volume Hamiltonian
\begin{equation}
\frac{d}{da}\mathcal{H}^{a,J}=\sum_{r=0}^{\infty}\sum_{j=1}^{J}\left[\mathcal{Y}^{J},\tilde{\mathcal{X}}_{j}^{a,(J,r)}\right]=\left[\mathcal{Y}^{J},\mathcal{H}^{a,J}\right].
\end{equation}
Clearly, the solution is
\begin{equation}
\mathcal{H}^{a,J}=\exp(a\mathcal{\mathcal{Y}}^{J})\mathcal{H}^{0,J}\exp(-a\mathcal{\mathcal{Y}}^{J}).
\end{equation}
We can see that the spectrum of the Hamiltonian is independent from
the rotations (parameters $\epsilon_{n}(\lambda)$) even for the finite
volume Hamiltonians!

\section{Lax-operators of the GL(N) long range spin chain}

In this section we demonstrate that, the Lax operators exist for the
$GL(N)$ long range spin chains. Now, we follow a slightly different
method than that described in the first section. At first, we fix
the integrable charge densities $\bar{q}_{k}^{2,\ell}$ which define
the charges
\begin{equation}
\bar{\mathcal{Q}}_{k}^{(\ell)}=\sum_{r=0}^{\ell}\lambda^{r}\bar{Q}_{k}^{(r)},\qquad\bar{Q}_{k}^{(\ell)}=\sum_{j=1}^{J}\bar{q}_{j}^{k,\ell}
\end{equation}
in the asymptotic region $J\geq\ell+k$. We derive the corresponding
Lax-operators from the commutation relations
\begin{equation}
\left[\bar{\mathcal{Q}}_{2}^{(\ell)},\check{\mathcal{T}}^{(\ell)}(u)\right]=\mathcal{O}(\lambda^{\ell+1}).\label{eq:comm}
\end{equation}
These relations are equivalent with the equations
\begin{equation}
\sum_{k=0}^{\ell}\left[\bar{Q}_{2}^{(k)},\check{T}^{(\ell-k)}(u)\right]=0.\label{eq:QT}
\end{equation}

We know that, the long range charges have $\gamma$-ambiguities i.e.
the linear combinations of the charges $\bar{\mathcal{Q}}_{k}^{(\ell)}$
define equivalent long range models. Since we only require the commutativity
(\ref{eq:comm}), the charges coming from the transfer matrix can
be in different $\gamma$ convention. Let us introduce the charges
from the derivatives of the transfer matrix
\begin{equation}
\mathcal{Q}_{k+1}^{(\ell)}=\frac{\partial^{k}}{\partial u^{k}}\log\check{\mathcal{T}}^{(\ell)}(u)\Biggr|_{u=0}+\mathcal{O}(\lambda^{\ell+1}).
\end{equation}
Since (\ref{eq:comm}) is satisfied, the new charges $\mathcal{Q}_{k+1}^{(\ell)}$
can be expressed by the original ones as \citep{Beisert:2005wv}
\begin{equation}
\mathcal{Q}_{r}^{(\ell)}=\gamma_{r,0}(\lambda)\mathbf{1}+\sum_{s=2}^{\ell+r}\gamma_{r,s}(\lambda)\bar{\mathcal{Q}}_{s}^{(\ell)}+\mathcal{O}(\lambda^{\ell+1}),
\end{equation}
where
\begin{equation}
\gamma_{r,s}(\lambda)=\sum_{k=\max(s-r,0)}^{\infty}\lambda^{k}\gamma_{r,s}^{(k)}.
\end{equation}
Substituting back, we can obtain an equivalent form
\begin{equation}
Q_{r}^{(\ell)}=\gamma_{r,0}^{(\ell)}\mathbf{1}+\sum_{s=2}^{\ell+r}\sum_{k=\max(s-r,0)}^{\ell}\gamma_{r,s}^{(k)}\bar{Q}_{s}^{(\ell-k)}.\label{eq:gamma}
\end{equation}
Let us write down the first three order of the Hamiltonian explicitly
\begin{align}
Q_{2}^{(0)} & =\gamma_{2,0}^{(0)}\mathbf{1}+\gamma_{2,2}^{(0)}\bar{Q}_{s}^{(0)},\\
Q_{2}^{(1)} & =\gamma_{2,0}^{(1)}\mathbf{1}+\gamma_{2,2}^{(0)}\bar{Q}_{2}^{(1)}+\gamma_{2,2}^{(1)}\bar{Q}_{2}^{(0)}+\gamma_{2,3}^{(1)}\bar{Q}_{3}^{(0)},\\
Q_{2}^{(2)} & =\gamma_{2,0}^{(2)}\mathbf{1}+\gamma_{2,2}^{(0)}\bar{Q}_{2}^{(2)}+\gamma_{2,2}^{(1)}\bar{Q}_{2}^{(1)}+\gamma_{2,2}^{(2)}\bar{Q}_{2}^{(0)}+\gamma_{2,3}^{(1)}\bar{Q}_{3}^{(1)}+\gamma_{2,3}^{(2)}\bar{Q}_{3}^{(0)}+\gamma_{2,4}^{(2)}\bar{Q}_{4}^{(0)}.
\end{align}

Our strategy is summarized on the right figure of \ref{fig:req}.
Assuming that we have the operators $\check{L}_{1}^{(r)},\check{J}_{1}^{(r)},\check{R}_{1}^{(\ell-1,r)}$
for $r=0,\dots,\ell-1$, we can calculate the transfer matrices $T^{(r)}(u)$
for $r=0,\dots,\ell-1$. Having an ansatz for $\check{L}_{1}^{(\ell)}$
we can also calculate $T^{(\ell)}(u)$, and using equation (\ref{eq:QT}),
we can obtain the explicit form of $\check{L}_{1}^{(\ell)}$. From
the equations (\ref{eq:Req}), we can calculate $\check{R}_{1}^{(\ell,r)}$
for $r=0,\dots,\ell-1$. After that, we can obtain $\check{R}_{1}^{(\ell,\ell)}$
from (\ref{eq:RLLeq}). Finally we can calculate $\check{J}_{1}^{(\ell)}$
from (\ref{eq:Jeq}) therefore we just have the next level operators
$\check{L}_{1}^{(r)},\check{J}_{1}^{(r)},\check{R}_{1}^{(\ell,r)}$
for $r=0,\dots,\ell$.

In this work we only calculate the first two order. We use the following
conventions for the previously calculated integrable charges \citep{Beisert:2005wv}
\begin{align}
\bar{q}_{1}^{2,0} & =()-(12),\label{eq:q20}\\
\bar{q}_{1}^{2,1} & =\alpha_{0}^{(1)}(-3()+4(12)-(13)),\label{eq:q21}\\
\bar{q}_{1}^{2,2} & =\left(\alpha_{0}^{(1)}\right)^{2}(20()-29(12)+10(13)-(14)-(1234)+(1243)+(1342)-(1432))+\nonumber \\
 & +\alpha_{0}^{(2)}(-3()+4(12)-(13))+\nonumber \\
 & +\frac{i}{2}\alpha_{1}^{(2)}(6(123)-(124)-6(132)-(134)+(142)+(143))+\nonumber \\
 & +\frac{1}{2}\beta_{2,3}^{(2)}(-4()+8(12)-2(123)+(124)-2(132)+(134)+(142)+(143)-\nonumber \\
 & \qquad\qquad-2(1234)+2(1243)+2(1342)-2(1432)-2(12)(34)-2(13)(24)).\label{eq:q22}
\end{align}
where $(j_{1}\dots j_{k})$ denotes the cycles of the permutation
group which act on the Hilbert space as 
\begin{equation}
(j_{1}\dots j_{k})\equiv P_{j_{1},j_{2}}P_{j_{2},j_{3}}\dots P_{j_{k-1},j_{k}},
\end{equation}
and the empty cycle is just the identity $()\equiv\mathbf{1}$. In
(\ref{eq:q20})-(\ref{eq:q22}) we already fixed the unphysical parameters
$\gamma_{r,s}$ and $\epsilon_{k}$. Our goal is to find the Lax-operator
$\check{\mathcal{L}}^{(2)}$which defines the transfer matrix which
generates a class of conserved charges corresponding to the physical
parameters $\alpha_{0}^{(1)},\alpha_{0}^{(2)},\alpha_{1}^{(2)},\beta_{2,3}^{(2)}$.
We will also need the leading order of the higher charges

\begin{align}
\bar{q}_{1}^{3,0} & =\frac{i}{2}\left((132)-(123)\right),\\
\bar{q}_{1}^{4,0} & =-(12)+(13)-\frac{1}{2}(1234)+\frac{1}{2}(1243)+\frac{1}{2}(1342)-\frac{1}{2}(1432).
\end{align}

\subsection*{Order 0}

In the zeroth order we have the usual nearest neighbor $GL(N)$ spin
chain which has the following Lax operator
\begin{equation}
\check{L}_{1}^{(0)}(u)=()-(12)u.
\end{equation}
We also have the zeroth order $R$- and $J$-operators as
\begin{equation}
\check{R}_{1}^{(0,0)}=\check{L}_{1}^{(0)}(u-v),\qquad\check{J}_{1}^{(0)}(u)=\frac{1}{1-u^{2}}\check{L}_{1}^{(0)}(-u).
\end{equation}
This Lax operator satisfies the regularity condition $\check{L}_{1}^{(0)}(0)=\mathbf{1}$
and the first derivative is
\begin{equation}
\frac{d}{du}\check{L}_{1}^{(0)}(u)\Biggr|_{u=0}=-(12)=-()+\bar{q}_{1}^{2,0},
\end{equation}
therefore we obtained the following $\gamma$ parameters
\begin{equation}
\gamma_{2,0}^{(0)}=-1,\quad\gamma_{2,2}^{(0)}=1.
\end{equation}

\subsection*{Order 1}

Having a general ansatz for $\check{L}_{1}^{(1)}(u)$ as
\[
\check{L}_{1}^{(1)}(u)=x_{1}(u)()+x_{2}(u)(12)+x_{3}(u)(123)+x_{4}(u)(132)+x_{5}(u)(13),
\]
the equation (\ref{eq:QT}) for $\ell=1$ fixes only two unknown functions
\[
x_{4}(u)=-x_{3}(u)-\frac{\alpha_{0}^{(1)}u^{2}}{u^{2}-1},\qquad x_{5}(u)=\frac{\alpha_{0}^{(1)}u}{u^{2}-1}.
\]
We can see that, the Lax operator contains three unfixed functions
which correspond to the $\gamma$-ambiguities of the definition of
charges. For the simplicity, let us fix these ambiguities as $x_{1}(u)=x_{2}(u)=x_{3}(u)=0$
therefore the Lax-operator simplifies as
\begin{equation}
\check{L}_{1}^{(1)}(u)=\frac{\alpha_{0}^{(1)}u}{u^{2}-1}(13)-\frac{\alpha_{0}^{(1)}u^{2}}{u^{2}-1}(132).
\end{equation}

The first order $R$-matrix $\check{R}^{(1,1)}$ can be obtained from
the equation (\ref{eq:RLLeq}):
\begin{align}
\check{R}^{(1,1)} & =-\frac{(u-v)(3u^{2}v^{2}-u^{2}-v^{2}-1)\alpha_{0}^{(1)}}{(u^{2}-1)(v^{2}-1)^{2}}(13)-\frac{u^{2}(u-v)\alpha_{0}^{(1)}}{(u^{2}-1)(v^{2}-1)}(1243)\nonumber \\
 & -\frac{v^{2}(u-v)\alpha_{0}^{(1)}}{(v^{2}-1)^{2}}(1342)+\frac{u(u-v)\alpha_{0}^{(1)}}{(u^{2}-1)(v^{2}-1)}(124)\nonumber \\
 & +\frac{2u(u-v)\alpha_{0}^{(1)}}{(u^{2}-1)(v^{2}-1)}(132)+\frac{u(u-v)\alpha_{0}^{(1)}}{(u^{2}-1)(v^{2}-1)}(243)\\
 & +\frac{uv(u-v)\alpha_{0}^{(1)}}{(u^{2}-1)(v^{2}-1)}(1324)+\frac{uv(u-v)\alpha_{0}^{(1)}}{(v^{2}-1)^{2}}(1423)\nonumber \\
 & +\frac{uv(u-v)^{2}(uv+1)\alpha_{0}^{(1)}}{(u^{2}-1)(v^{2}-1)^{2}}(13)(24)-\frac{(u-v)(u^{2}v^{2}-1)\alpha_{0}^{(1)}}{(u^{2}-1)(v^{2}-1)^{2}}(24)\nonumber \\
 & -\frac{2v(u-v)\alpha_{0}^{(1)}}{(v^{2}-1)^{2}}(123)-\frac{v(u-v)\alpha_{0}^{(1)}}{(v^{2}-1)^{2}}(142)-\frac{v(u-v)\alpha_{0}^{(1)}}{(v^{2}-1)^{2}}(234).\nonumber 
\end{align}
This Lax operator satisfies the regularity condition $\check{L}_{1}^{(1)}(0)=0$
and the first derivative is
\begin{equation}
\frac{d}{du}\check{L}_{1}^{(1)}(u)\Biggr|_{u=0}=\bar{q}_{1}^{2,1}+4\alpha_{0}^{(1)}\bar{q}_{1}^{2,0}-\alpha_{0}^{(1)}\mathbf{1},
\end{equation}
therefore the we obtained the following $\gamma$ parameters
\begin{equation}
\gamma_{2,0}^{(1)}=-\alpha_{0}^{(1)},\quad\gamma_{2,2}^{(1)}=4\alpha_{0}^{(1)},\quad\gamma_{2,3}^{(1)}=0.
\end{equation}

\subsection*{Order 2}

The following Lax operator
\begin{align}
\check{L}_{1}^{(2)}(u) & =y_{1}(u)(142)+y_{2}(u)(14)+y_{3}(u)(132)+y_{4}(u)(13)\nonumber \\
 & +y_{5}(u)(124)+y_{6}(u)(143)+y_{7}(u)(134)+y_{8}(u)(1432)\nonumber \\
 & +y_{9}(u)(1324)+y_{10}(u)(1423)+y_{11}(u)(1243)\nonumber \\
 & +y_{12}(u)(12)(34)+y_{13}(u)(1234)+y_{14}(u)(14)(23)+y_{15}(u)(13)(24)
\end{align}
is a solution of the equation (\ref{eq:QT}) for $\ell=2$, where
\begin{align}
y_{1}(u) & =\frac{-2u^{2}(u^{2}-1)\left(\alpha_{0}^{(1)}\right)^{2}+u\beta_{2,3}^{(2)}+iu\alpha_{1}^{(2)}}{2(u^{2}-1)^{2}}, & y_{9}(u) & =\frac{u^{2}\beta_{2,3}^{(2)}+iu^{2}\alpha_{1}^{(2)}}{2(u^{2}-1)^{2}},\qquad y_{12}(u)=\frac{u\beta_{2,3}^{(2)}}{u^{2}-1},\\
y_{2}(u) & =\frac{2u(u^{2}-1)\left(\alpha_{0}^{(1)}\right)^{2}-u^{2}\beta_{2,3}^{(2)}-iu^{2}\alpha_{1}^{(2)}}{2(u^{2}-1)^{2}}, & y_{10}(u) & =\frac{u^{2}\beta_{2,3}^{(2)}+iu^{2}\alpha_{1}^{(2)}}{2(u^{2}-1)^{2}},\\
y_{5}(u) & =\frac{-u\beta_{2,3}^{(2)}+iu\alpha_{1}^{(2)}}{2(u^{2}-1)}, & y_{11}(u) & =\frac{u^{2}\beta_{2,3}^{(2)}+iu^{2}(2u^{2}-1)\alpha_{1}^{(2)}}{2(u^{2}-1)^{2}},\\
y_{6}(u) & =\frac{-u\beta_{2,3}^{(2)}-iu\alpha_{1}^{(2)}}{2(u^{2}-1)}, & y_{13}(u) & =\frac{u^{2}(u^{2}-2)\beta_{2,3}^{(2)}-iu^{4}\alpha_{1}^{(2)}}{2(u^{2}-1)^{2}},\\
y_{7}(u) & =\frac{u\beta_{2,3}^{(2)}+iu(2u^{2}-1)\alpha_{1}^{(2)}}{2(u^{2}-1)^{2}}, & y_{14}(u) & =\frac{-u^{3}\beta_{2,3}^{(2)}-iu^{3}\alpha_{1}^{(2)}}{2(u^{2}-1)^{2}},\\
y_{8}(u) & =\frac{-u^{2}\beta_{2,3}^{(2)}-iu^{2}\alpha_{1}^{(2)}}{2(u^{2}-1)^{2}}, & y_{15}(u) & =\frac{u(u^{2}-2)\beta_{2,3}^{(2)}-iu^{3}\alpha_{1}^{(2)}}{2(u^{2}-1)^{2}},
\end{align}
and
\begin{align}
y_{3}(u) & =\frac{2u^{2}(2u^{2}-5)\left(\alpha_{0}^{(1)}\right)^{2}+u(-2u^{3}+3u^{2}+2u-4)\beta_{2,3}^{(2)}-2u^{2}(u^{2}-1)\alpha_{0}^{(2)}+iu^{3}(-2u^{2}+5)\alpha_{1}^{(2)}}{2(u^{2}-1)^{2}},\\
y_{4}(u) & =\frac{u(-4u^{2}+10)\left(\alpha_{0}^{(1)}\right)^{2}-u(u^{3}-2u^{2}-2u+2)\beta_{2,3}^{(2)}+2u(u^{2}-1)\alpha_{0}^{(2)}+iu^{2}(u^{2}-4)\alpha_{1}^{(2)}}{2(u^{2}-1)^{2}}.
\end{align}
 The $R$-matrix $\check{R}^{(2,2)}$ (which satisfies the relation
(\ref{eq:RLLeq})) also exists but it has very complicated form. It
can be written as a linear combination of 375 cycles therefore we
omit to describe the explicit form.

This Lax operator satisfies the regularity condition $\check{L}_{1}^{(2)}(0)=0$
and the first derivative is
\begin{align}
\frac{d}{du}\check{L}_{1}^{(2)}(u)\Biggr|_{u=0}= & \bar{q}_{1}^{2,2}+(2i\beta_{2,3}^{(2)}+6\alpha_{1}^{(2)})\bar{q}_{1}^{3,0}-(\left(\alpha_{0}^{(1)}\right)^{2}+\beta_{2,3}^{(2)})\bar{q}_{1}^{4,0}+\nonumber \\
 & +4\alpha_{0}^{(1)}\bar{q}_{1}^{2,1}+(-12\left(\alpha_{0}^{(1)}\right)^{2}+5\beta_{2,3}^{(2)}+4\alpha_{0}^{(2)})\bar{q}_{1}^{2,0}-\nonumber \\
 & +(4\left(\alpha_{0}^{(1)}\right)^{2}-3\beta_{2,3}^{(2)}-\alpha_{0}^{(2)})\mathbf{1}.
\end{align}
therefore we obtained the following $\gamma$ parameters
\begin{equation}
\begin{split}\gamma_{2,0}^{(2)}= & 4\left(\alpha_{0}^{(1)}\right)^{2}-3\beta_{2,3}^{(2)}-\alpha_{0}^{(2)},\\
\gamma_{2,2}^{(2)}= & -12\left(\alpha_{0}^{(1)}\right)^{2}+5\beta_{2,3}^{(2)}+4\alpha_{0}^{(2)},\\
\gamma_{2,3}^{(2)}= & 2i\beta_{2,3}^{(2)}+6\alpha_{1}^{(2)},\\
\gamma_{2,4}^{(2)}= & -\left(\alpha_{0}^{(1)}\right)^{2}-\beta_{2,3}^{(2)}.
\end{split}
\end{equation}

\end{document}